\documentclass{article}

\usepackage[final]{neurips_2024}

\usepackage[utf8]{inputenc} 
\usepackage[T1]{fontenc}    
\usepackage{hyperref}       
\usepackage{url}            
\usepackage{booktabs}       
\usepackage{amsfonts}       
\usepackage{nicefrac}       
\usepackage{microtype}      
\usepackage{xcolor}         
\usepackage{graphicx}
\usepackage{subcaption}
\usepackage{amsmath}
\usepackage{graphicx}
\usepackage{tikz}
\usetikzlibrary{positioning}
\usetikzlibrary{calc} 
\usepackage{pgfplots}
\pgfplotsset{compat=1.18} 
\usetikzlibrary{pgfplots.statistics}
\usetikzlibrary{pgfplots.groupplots}
\usepgfplotslibrary{colorbrewer}
\usepackage{placeins}

\usepackage{natbib}
\setcitestyle{square,numbers}

\title{Conditional Hallucinations for Image Compression}

\author{%
    Till Aczel \\ ETH Zürich \\ \texttt{taczel@ethz.ch} \\ \And
    Roger Wattenhofer \\ ETH Zürich \\ \texttt{wattenhofer@ethz.ch} \\
}

\begin{document}

\maketitle

\begin{abstract}

In lossy image compression, models face the challenge of either hallucinating details or generating out-of-distribution samples due to the information bottleneck.
This implies that at times, introducing hallucinations is necessary to generate in-distribution samples.
The optimal level of hallucination varies depending on image content, as humans are sensitive to small changes that alter the semantic meaning.
We propose a novel compression method that dynamically balances the degree of hallucination based on content.
We collect data and train a model to predict user preferences on hallucinations.
By using this prediction to adjust the perceptual weight in the reconstruction loss, we develop a \textbf{Con}ditionally \textbf{Ha}llucinating compression model (\textbf{ConHa}) that outperforms state-of-the-art image compression methods. 
Code and images are available at \href{https://polybox.ethz.ch/index.php/s/owS1k5JYs4KD4TA}{https://polybox.ethz.ch/index.php/s/owS1k5JYs4KD4TA}.

  
\end{abstract}

\section{Introduction}
Lossy compression is characterized by a trade-off between reducing the number of communicated bits (rate) and the undesirable introduction of content changes (distortion), presenting a constrained optimization challenge.
For image compression measuring the distortion or ``how close a compressed image is to the original image'' is a challenging task.
A distortion metric aligned with human perception should assess not only pixel-wise similarity but also how much content changes and the overall realism of the reconstructed image.
Applying a slight blur to an image containing small text can make it unreadable, drastically increasing distortion, while blurring an image of a uniform blue sky has minimal impact.
For texture-like content, such as grass, freckles, and stone walls, generating pixels that realistically match a given texture is more important than reconstructing precise pixel values; 
generating any sample from the distribution of a texture is generally sufficient.
Conversely, for images featuring small text, straight lines, and small faces, data reconstruction should be pixel-by-pixel to preserve fidelity \cite{mentzer2020high}.
Distortion metrics often struggle to capture these subtleties of human perceptual preference.

\begin{figure}[!ht]
    \centering
    \begin{tabular}{ccc}
    \rotatebox{90}{\textbf{ \hspace{1.5em} Original}} & \includegraphics[width=0.45\linewidth, trim=187.5px 390px 657.5px 40px, clip]{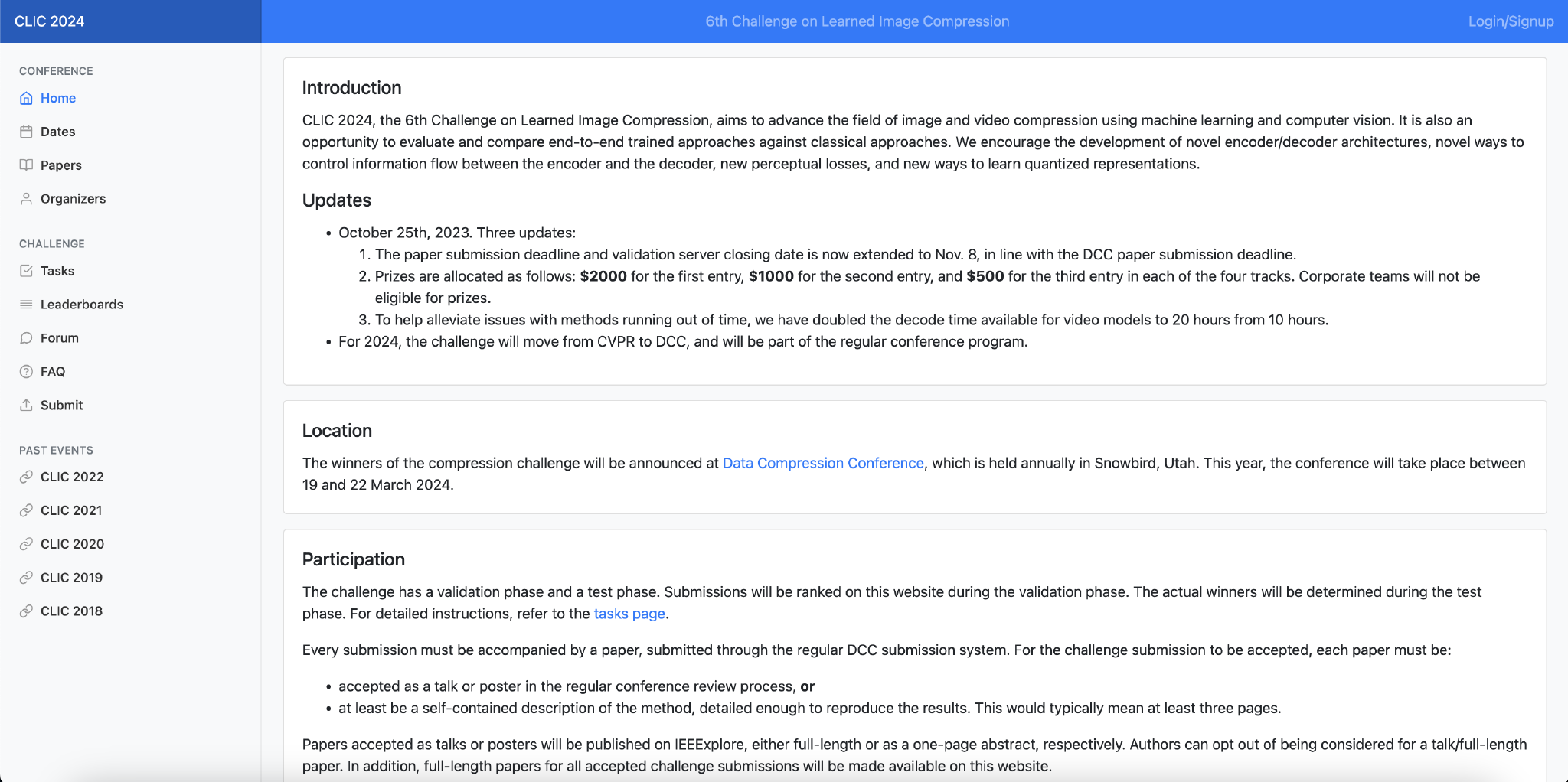}  & \includegraphics[width=0.45\linewidth, trim=0px 700px 1655px 0px, clip]{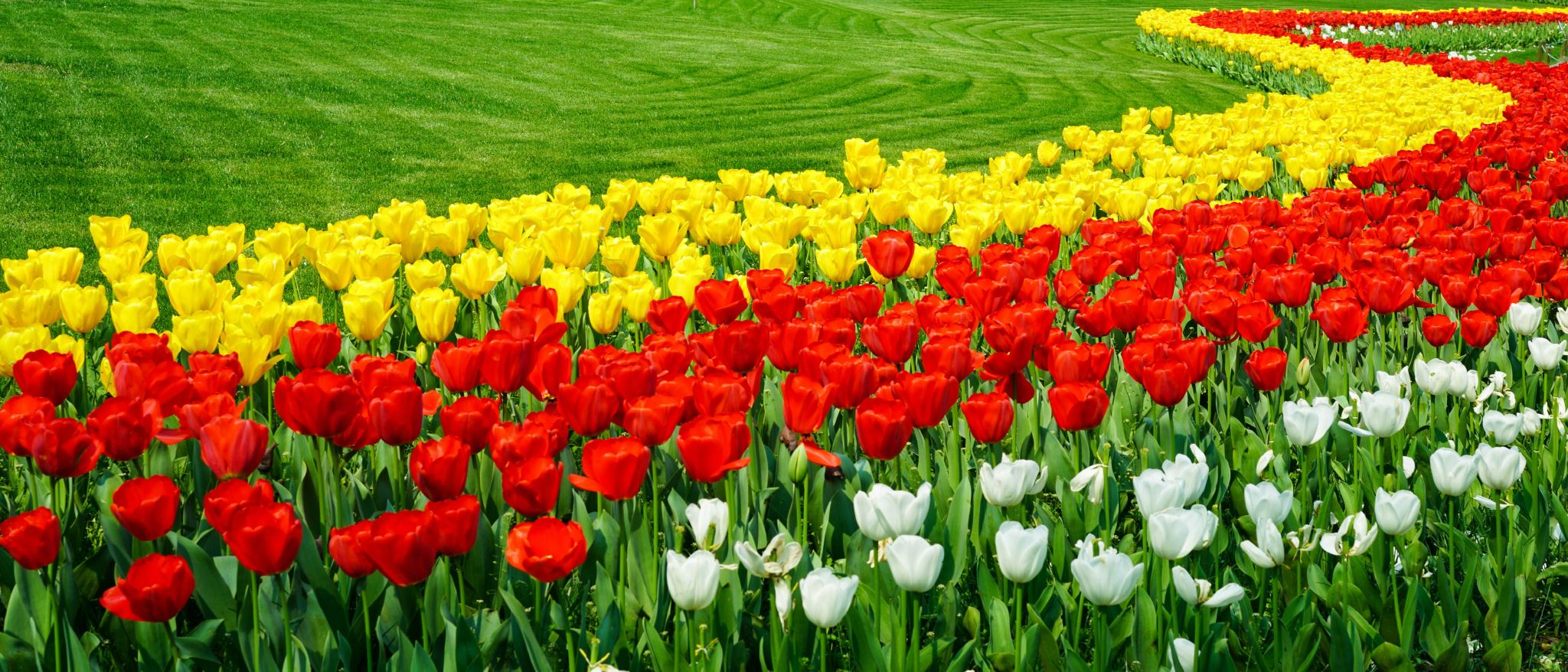} \\
        
    \rotatebox{90}{\textbf{\hspace{0.1em} No hallucinations}}& \includegraphics[width=0.45\linewidth, trim=375px 780px 1315px 80px, clip]{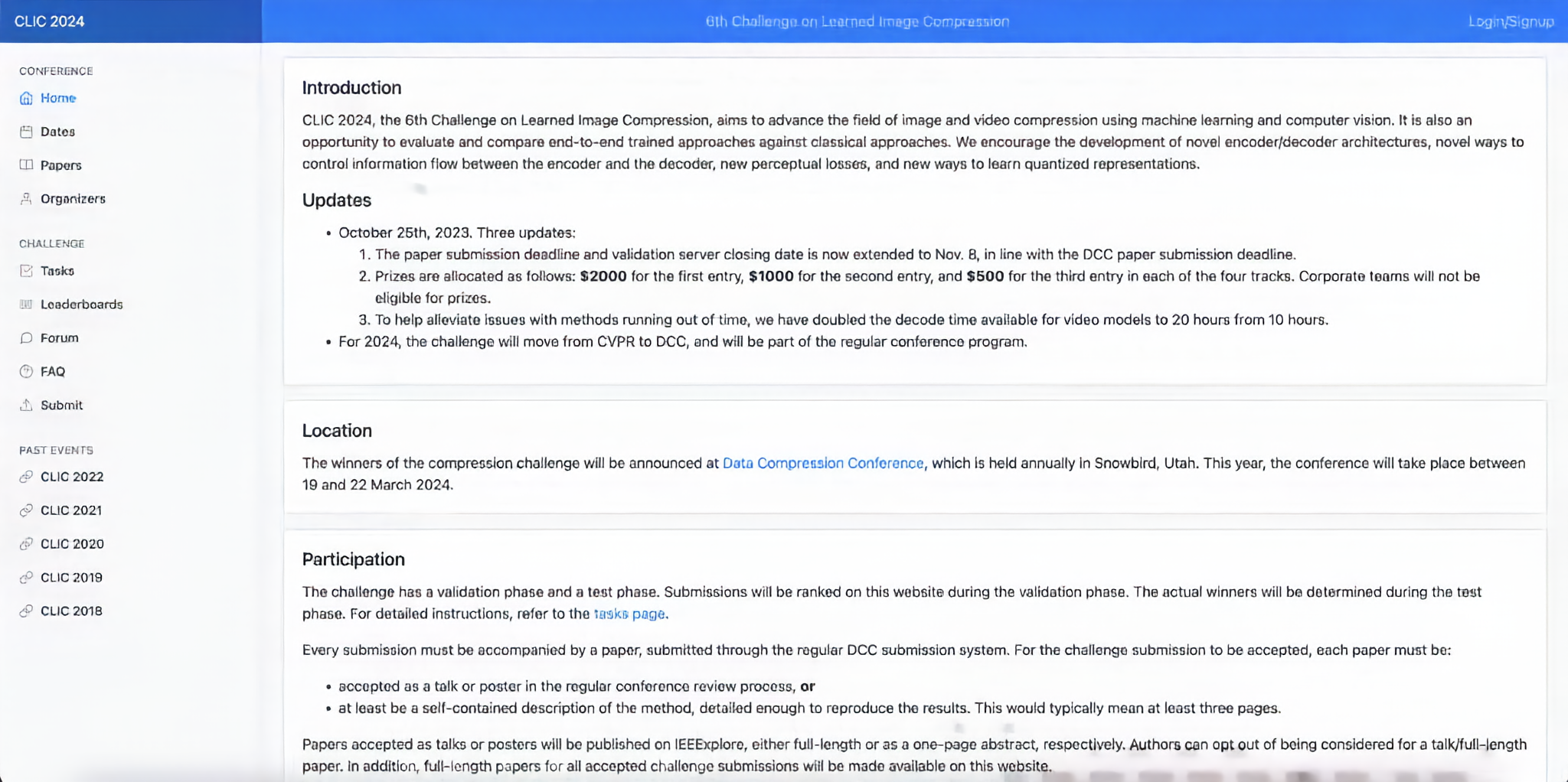} &  \includegraphics[width=0.45\linewidth, trim=0px 700px 1655px 0px, clip]{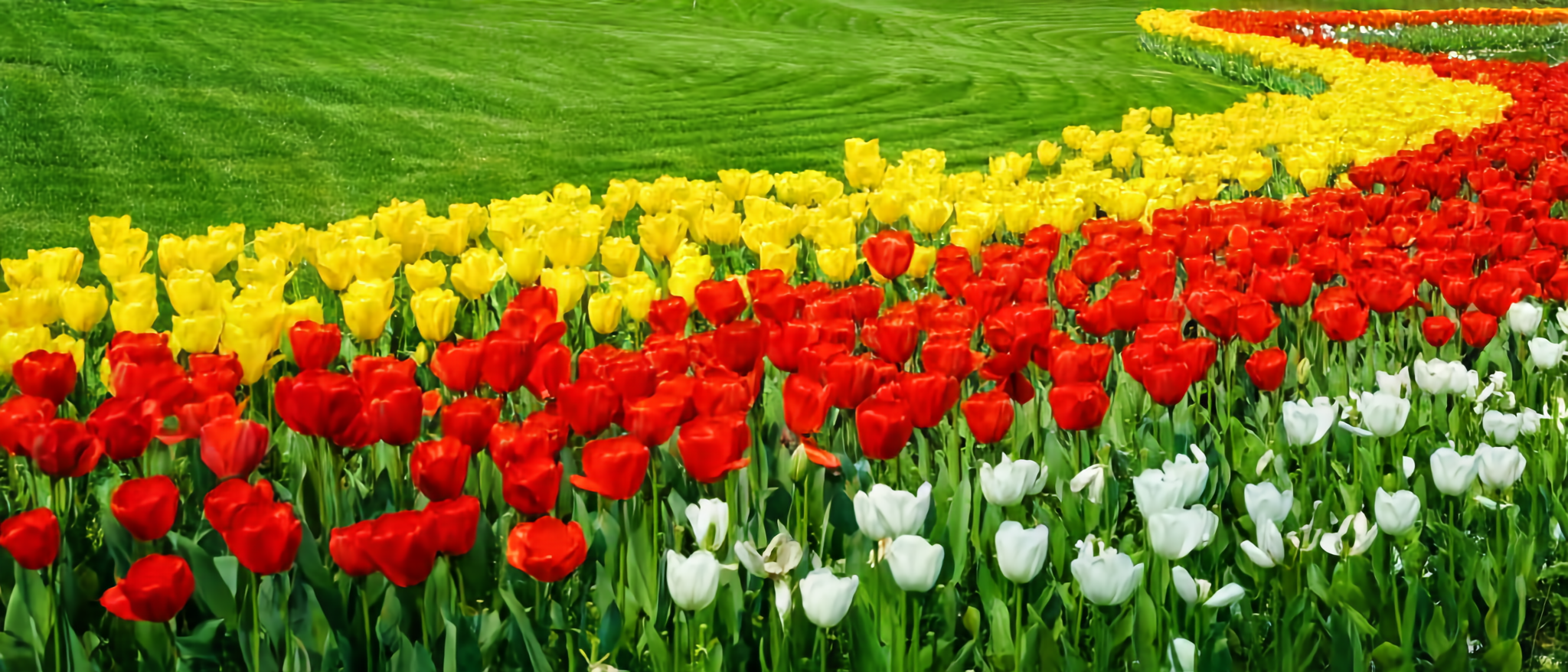} \\


    \rotatebox{90}{\textbf{\hspace{0.7em} In distribution}} & \includegraphics[width=0.45\linewidth, trim=375px 780px 1315px 80px, clip]{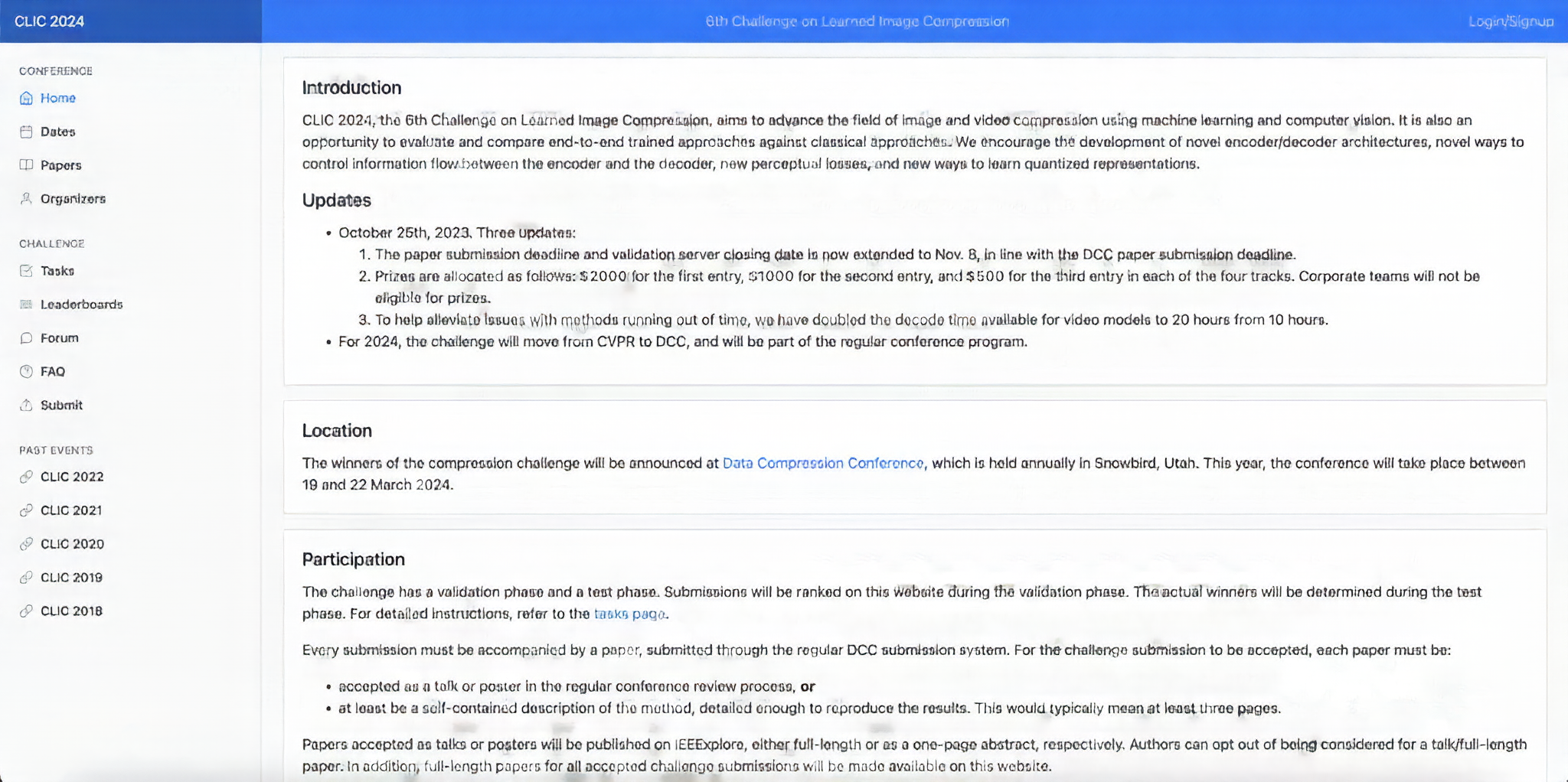} & \includegraphics[width=0.45\linewidth, trim=0px 700px 1655px 0px, clip]{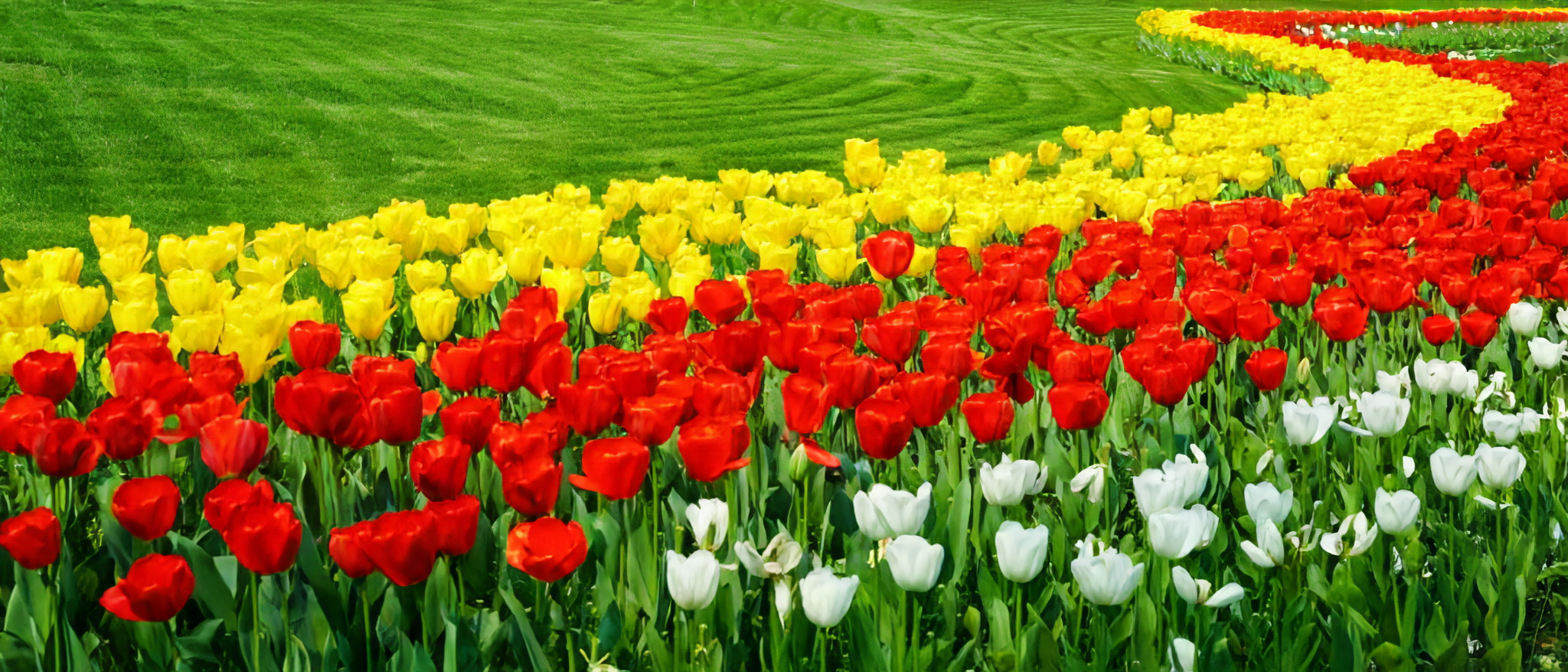} \\

    \end{tabular}
\caption{
1) Original image 
2) compressed with an MSE distortion optimized model 
3) compressed with a hallucinating MSE+GAN optimized model. 
For the left image containing text, hallucinations degrade the image quality. 
For the right image containing grass, an in-distribution sample with hallucinations produces a higher perceptual quality.}
\label{fig:HD}
\end{figure}

Learned image compression methods directly optimize the rate-distortion trade-off.
Thus the limitations of distortion metrics do not only impact their evaluation but also affect the training process.
To overcome these shortcomings, state-of-the-art methods \cite{mentzer2020high, he2022po} optimize the rate-distortion-perception trade-off, rather than just the rate-distortion trade-off. 
Here, perception refers to how closely the compressed image resembles the original in terms of realness or visual fidelity.
This can be achieved with a GAN-like discriminator, encouraging models to generate samples that match the distribution of real images, enhancing perceptual quality.
However, optimizing this trade-off can introduce hallucinations, where generated images contain details not present in the original data.
This is undesirable, particularly when small hallucinations alter the semantic content.
In Figure \ref{fig:HD}, we present two images: one with small text and another with grass, compressed by two models—one avoiding hallucinations and the other generating in-distribution samples.
The optimal rate-distortion-perception trade-off varies based on image content and compression rate.
For text, hallucinations should be avoided, while for grass, generating in-distribution samples replicating texture, even with hallucinations, is preferred.

In this paper, we gather labels and train a classifier to predict user preferences for the level of detail to hallucinate, based on the original uncompressed image.
By using the classifier's prediction as a weight for the GAN discriminator in the reconstruction loss, we train a \textbf{Con}ditionally \textbf{Ha}llucinating compression model, \textbf{ConHa}.
This model automatically adjusts the reconstruction process based on image content, determining whether to minimize hallucinations or generate in-distribution samples.
To the best of our knowledge, this is the first work to introduce automatic balancing between distortion and perception in image compression.
The degree of hallucination varies not only between different images but also within a single image.
By dynamically adjusting hallucination levels, we surpass state-of-the-art compression methods.




\section{Related work}

Traditional lossy image compression models like JPEG \cite{wallace1991jpeg}, JPEG2000 \cite{taubman2002jpeg2000}, WebP \cite{googlewebp}, and BPG \cite{bpg} are widely used.
Learned lossy image compression, which optimizes the relaxed rate-distortion trade-off, is conceptually similar to Variational Autoencoders (VAEs) \cite{kingma2013auto, rezende2014stochastic}, but requires latent quantization for compression \cite{balle2016end}.
Using this similarity, \cite{balle2016end} developed a non-linear codec that outperforms traditional methods.
Building on this, they introduced Hyperprior \cite{balle2018variational}, a hierarchical VAE that jointly learns and compresses both the latent variable and its prior.
More recently, ELIC \cite{he2022elic} explores architectural improvements to propose a high-performing and computationally efficient learned image codec.

In the rate-distortion trade-off, the rate has an accurate approximation, but measuring distortion is much harder.
Handcrafted metrics like PSNR and MS-SSIM often poorly align with user studies and can even negatively correlate with human preferences among the best-performing models \cite{compressioncc}.
The hidden representations of trained CNNs correlate strongly with human preferences \cite{zhang2018unreasonable}, which the Learned Perceptual Image Patch Similarity (LPIPS) metric \cite{zhang2018unreasonable} leverages by fine-tuning a CNN on two-alternative forced choice (2AFC) data.
Distortion metrics often fail to accurately capture human perception, leaving models vulnerable to adversarial attacks and resulting in out-of-distribution samples.
The Fréchet Inception Distance (FID) \cite{heusel2017gans} measures alignment between image distributions, evaluating the realism of generated images.
Generative Adversarial Networks (GANs) \cite{goodfellow2014generative} address this by using a generator and a discriminator to ensure in-distribution sample generation.
Variants of GAN-like adversarial losses \cite{rippel2017real, santurkar2018generative, tschannen2018deep, agustsson2019generative, blau2019rethinking, mentzer2020high, he2022po} effectively rectify compression artifacts and generate desired distribution samples.
Notably, \cite{blau2019rethinking} introduces the rate-distortion-perception trade-off, using perception as a metric for realism assessment.
HiFiC \cite{mentzer2020high} is a state-of-the-art image compressor favored in user studies over previous codecs.

Attaining perfect realism often leads to undesired distortion \cite{zhang2021universal} and introduces hallucinations into generated content.
Our research aims to automatically determine the optimal balance between distortion and perception.
While some methods allow manual control over the distortion-perception trade-off \cite{agustsson2023multi, iwai2024controlling}, enabling users to set realism levels for the entire image, our approach automates this process.
Automation is beneficial as users typically prefer convenience and may not clearly understand their desired level of hallucinations.
Instead of relying on user input for adjustments, our method selects realism levels for individual image parts based on content, allowing users to enjoy in-distribution samples with minimized hallucinations without altering the semantic meaning.


\section{Methodology}

Human preferences in the trade-off between rate, distortion, and perception vary with image content.
A compression model reflecting human perception must balance avoiding hallucinations and maintaining fidelity to in-distribution samples.
To capture this nuance, we collected a dataset of 5,408 preference choices between a Mean Squared Error (MSE)-based model and a Generative Adversarial Network (GAN)-based model.
First, we learn a preference model that predicts which compression method is preferred based on image content.
The perceptual compression loss is then scaled according to this prediction.
For images where preserving exact content is crucial and hallucinations are undesirable, the compression loss emphasizes rate and distortion.
Conversely, for images where generating an in-distribution sample suffices and exact pixel-wise accuracy is less critical, the model optimizes for rate, distortion, and also perceptual quality.
This approach yields a compression model that aligns better with human perceptual preferences.

\subsection{Hallucination-distribution preference model} \label{sec:pref_model}

\begin{figure}[!tb]
    \centering
    \begin{tikzpicture}
        \node[anchor=south west,inner sep=0] (image) at (0,0) {\includegraphics[width=0.95\textwidth]{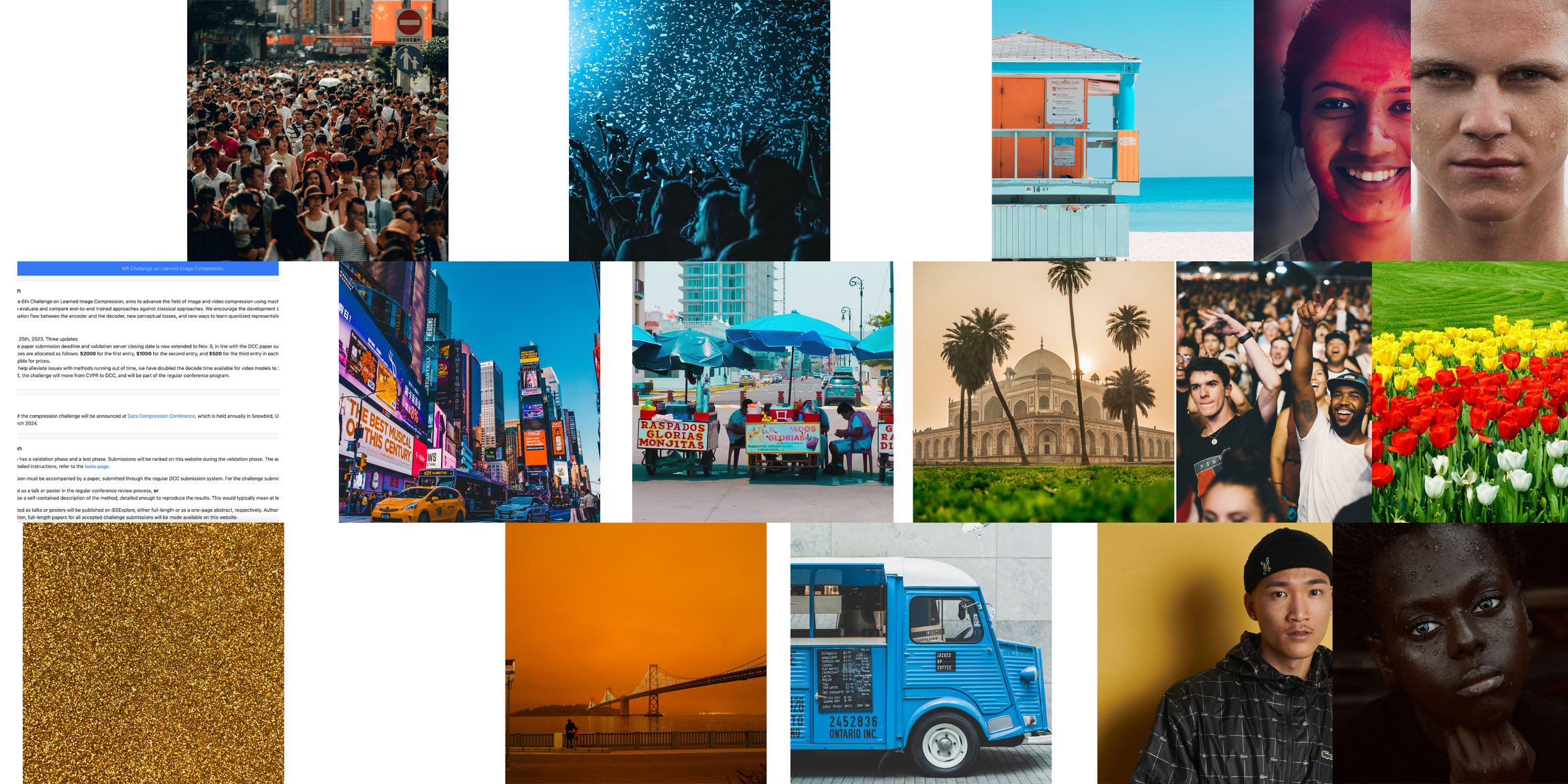}};
        \draw[<->, line width=1.pt] (0,-0.12) -- (0.95\textwidth,-0.12) node[pos=0.12, below] {No hallucinations} node[pos=0.9, below] {In distribution};
    \end{tikzpicture}
    \caption{Samples from the CLIC 2024 image test set with $w$ as their x coordinate.
    Images predicted to perform better without hallucinations are on the left, while those predicted to excel with in-distribution samples are on the right.
    }
    \label{fig:classifier_result_with_axes}
\end{figure}

To align the compression model's loss term with human preferences, obtaining human labels for all crops of all training images is too expensive.
Instead, we train a preference model $M_{P}$ on the labeled data, namely a binary classifier, which can then be utilized during the compression model training.
We provide further training details in Appendix \ref{app:training_details}.
In Figure \ref{fig:classifier_result_with_axes}, we visualize the preference model $M_{P}$ predictions.
Images with text, small faces, or straight lines (where hallucinations can alter semantic content) are on the left, while images of grass, skin, cloth, and other textures are on the right.
Unlike LPIPS, which has two inputs with variable distortion types, $M_{P}$ only needs to predict distortion-realism human preferenc.
This constraint allows the preference model to learn human preferences more accurately.


\subsection{Compression model}

\textbf{Rate-distortion optimized compression:}
We follow the autoencoder-based learned lossy image compression that optimizes the rate-distortion trade-off.
This method comprises a learned encoder $E$ and generator $G$. 
An image $x$ from distribution $p_x$ is encoded by $E$ into a quantized representation $y$, which is then decoded by $G$ to reconstruct the image $x'$.
The distribution of the quantized representation $y$ is learned by a probability model $P$, which is then used to further compress $y$ using an entropy coding algorithm.
The rate-distortion trade-off can be jointly optimized by:
\begin{equation}
    \mathcal{L}_{rd}=\mathbb{E}_{x \sim p_x} \left[ \lambda r(y) + d(x, x') \right],
\end{equation}
where $\lambda$ controls the trade-off, r is an approximation of the bit-stream length $r(y)=-log(P(y))$, and $d$ is a distortion metric (in our case MSE).

\textbf{Rate-distortion-perception optimized compression:} 
Models optimizing the rate-distortion trade-off with imperfect distortion metrics generate out-of-distribution samples with visible artifacts.
To address this, the trade-off can be extended to include perceptual quality, leading to a rate-distortion-perception trade-off.
However, current perceptual metrics like LPIPS are vulnerable to adversarial attacks, causing models to struggle with accurate image reconstruction.
Previous works \cite{rippel2017real, santurkar2018generative, tschannen2018deep, agustsson2019generative, blau2019rethinking, mentzer2020high, he2022po} improve the variational autoencoder by integrating a perceptual metric and a GAN-like discriminator, resulting in the objective:
\begin{equation} \label{equ:rdp}
    \mathcal{L}_{rdp}=\mathbb{E}_{x \sim p_x} \left[ \lambda r(y) + d(x, x') + \beta (d_{LPIPS}(x, x')-log(D(x',y)))\right],
\end{equation}
where $\beta$ is the perception weight, $d_{LPIPS}$ is the LPIPS metric and $D$ is a GAN-like discriminator.
The compression model and discriminator are trained alternately, with the discriminator optimizing:
\begin{equation}
    \mathcal{L}_{D}=\mathbb{E}_{x \sim p_x} \left[-log(1-D(x',y)))\right]+\mathbb{E}_{x \sim p_x} \left[-log(D(x,y)))\right]
\end{equation}
We have grouped  $D_{LPIPS}$ and $D$ together since these terms control the perceptual quality and enforce that the reconstructed image is within the image distribution $p_x$.

\textbf{Content-dependent rate-distortion-perception optimized compression:}
The optimal balance in the rate-distortion-perception trade-off varies with the image content.
By adjusting the weight assigned to perceptual quality in the rate-distortion-perception loss function, the compression model learns when to add details to create an in-distribution sample:
\begin{equation} \label{equ:wrdp}
    \mathcal{L}_{wrdp}=\mathbb{E}_{x \sim p_x} \left[ \lambda r(y) + d(x, x') + \beta w (d_{LPIPS}(x, x')-log(D(x',y)))\right],
\end{equation}
where $w$ is the weight predicted by the hallucination-distortion preference model $M_{P}$ for image $x$.
For images of at least 64 $\times$ 64, there is no information flow between distant pixels. 
Thus, while $w$ remains static during training, the model determines hallucination levels based on a limited context window during inference, resulting in varying amounts of hallucinations within an image.
Although we use HiFiC as our baseline, this correction method can be applied to any model utilizing a VAE with a GAN discriminator.
Training details are in Appendix \ref{app:training_details}.

\section{Experiments}

\textbf{Datasets:} 
Due to our two-stage training process, we require two distinct training sets. 
The preference model, is trained using the DIV2K training set \cite{div2k}, while the compression models are trained on the Vimeo90K dataset \cite{TOFlow}. 
The user study is performed on the CLIC 2024 image test set \cite{compressioncc}, which consists of 32 diverse high-resolution (above 2 megapixels) images.

\begin{figure}[!bt]
    \centering
    \begin{tikzpicture}
    \node (big_gan_ratio) at (0,-0.4) {\includegraphics[width=0.5\linewidth, trim=0 90px 1024px 400px, clip]{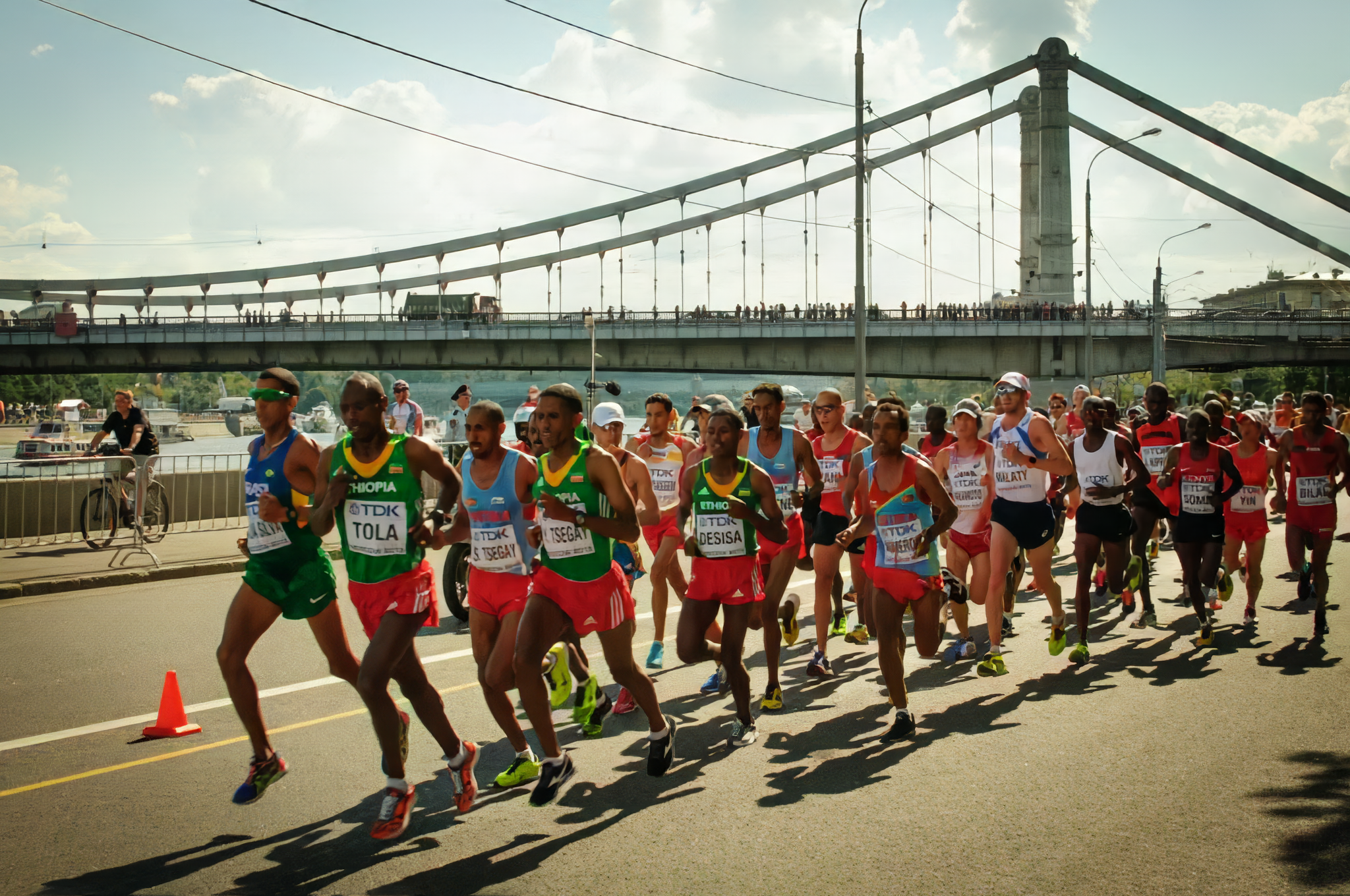}};
    \node (big_orig) at (0.5\linewidth,-0.4) {\includegraphics[width=0.5\linewidth, trim=1024px 90px 0 400px, clip]{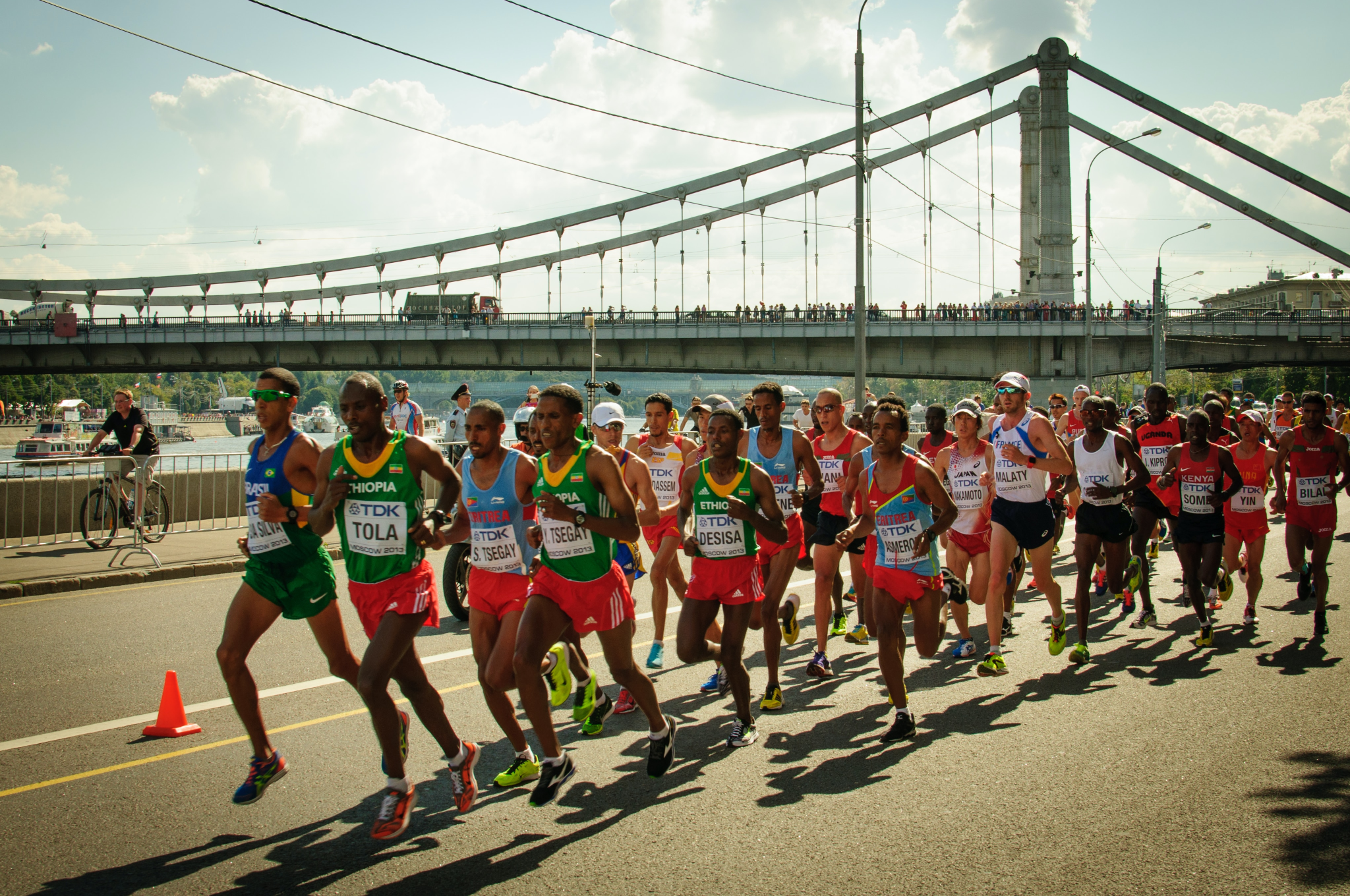}};    
    
    \node (ground_orig)      at (-0.083333\linewidth,-3.4)  {\includegraphics[width=0.333333\linewidth, trim=100px 228px 1720px 1050px, clip]{figures/qualitative/original.png}};
    \node (ground_mse)       at (-0.083333\linewidth,-5.06)  {\includegraphics[width=0.333333\linewidth, trim=100px 228px 1720px 1050px, clip]{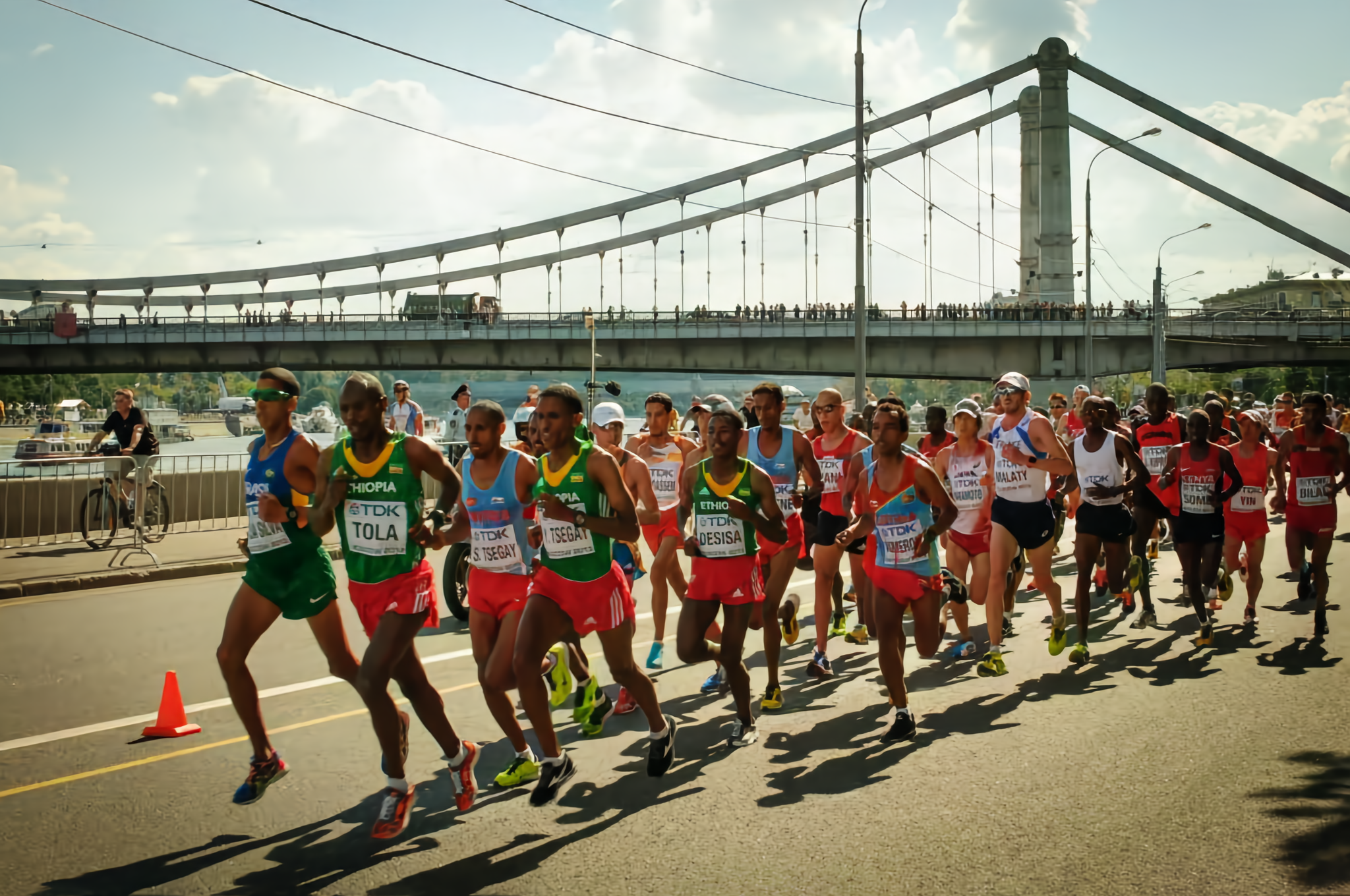}};
    \node (ground_ratio_gan) at (-0.083333\linewidth,-6.72)  {\includegraphics[width=0.333333\linewidth, trim=100px 228px 1720px 1050px, clip]{figures/qualitative/gan_ratio.png}};
    \node (ground_gan)       at (-0.083333\linewidth,-8.40) {\includegraphics[width=0.333333\linewidth, trim=100px 228px 1720px 1050px, clip]{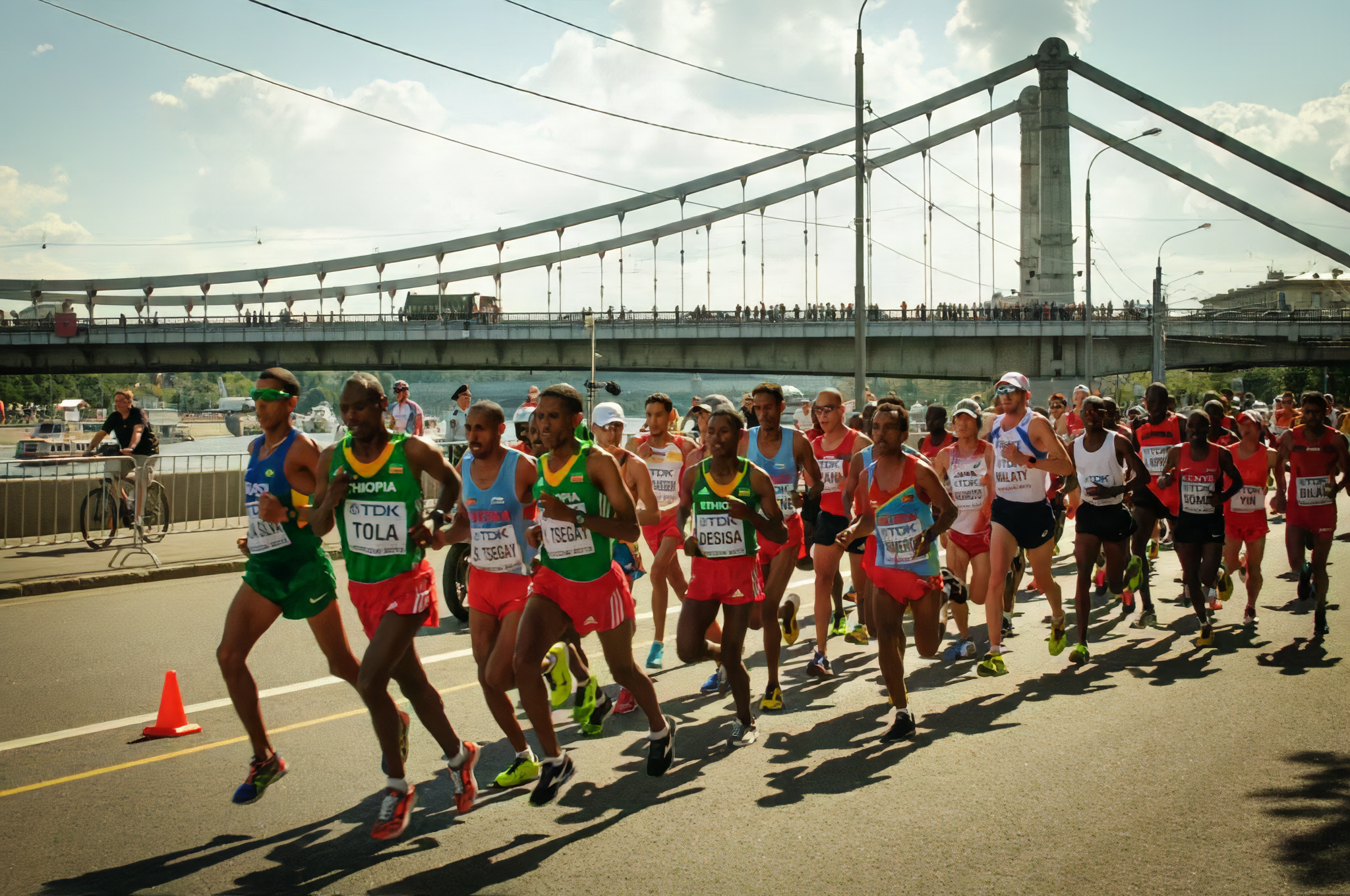}};
    
    \node (face_orig)        at (0.25\linewidth,-3.4)      {\includegraphics[width=0.333333\linewidth, trim=1550px 698px 270px 580px, clip]{figures/qualitative/original.png}};
    \node (face_mse)         at (0.25\linewidth,-5.06)      {\includegraphics[width=0.333333\linewidth, trim=1550px 698px 270px 580px, clip]{figures/qualitative/mse.png}};
    \node (face_ratio_gan)   at (0.25\linewidth,-6.72)      {\includegraphics[width=0.333333\linewidth, trim=1550px 698px 270px 580px, clip]{figures/qualitative/gan_ratio.png}};
    \node (face_gan)         at (0.25\linewidth,-8.40)     {\includegraphics[width=0.333333\linewidth, trim=1550px 698px 270px 580px, clip]{figures/qualitative/gan.png}};
    
    \node (truck_orig)       at (0.583333\linewidth,-3.4)  {\includegraphics[width=0.333333\linewidth, trim=560px 848px 1260px 430px, clip]{figures/qualitative/original.png}};
    \node (truck_mse)        at (0.583333\linewidth,-5.06)  {\includegraphics[width=0.333333\linewidth, trim=560px 848px 1260px 430px, clip]{figures/qualitative/mse.png}};
    \node (truck_ratio_gan)  at (0.583333\linewidth,-6.72)  {\includegraphics[width=0.333333\linewidth, trim=560px 848px 1260px 430px, clip]{figures/qualitative/gan_ratio.png}};
    \node (truck_gan)        at (0.583333\linewidth,-8.40) {\includegraphics[width=0.333333\linewidth, trim=560px 848px 1260px 430px, clip]{figures/qualitative/gan.png}};
    
    \node[fill=black!3, minimum width=0.501\linewidth] (input) at (0.\linewidth,2.8) {Original};
    \node[fill=black!3, minimum width=0.501\linewidth] (input) at (0.5\linewidth,2.8) {ConHa (ours): 0.169bpp};
    
    \node[fill=black!3, rotate=90, font=\sffamily, minimum width=1.69cm, minimum height=0.45cm] (input) at (-0.235\linewidth,-3.4) {\scriptsize Original};
    \node[fill=black!3, rotate=90, font=\sffamily, minimum width=1.69cm, minimum height=0.45cm] (input) at (-0.203\linewidth,-3.4) {\scriptsize };
    \node[fill=black!3, rotate=90, font=\sffamily, minimum width=1.69cm, minimum height=0.45cm] (input) at (-0.235\linewidth,-5.06) {\scriptsize Hyperprior};    
    \node[fill=black!3, rotate=90, font=\sffamily, minimum width=1.69cm, minimum height=0.45cm] (input) at (-0.203\linewidth,-5.06) {\scriptsize 0.165bpp};
    \node[fill=black!3, rotate=90, font=\sffamily, minimum width=1.69cm, minimum height=0.45cm] (input) at (-0.235\linewidth,-6.72)  {\scriptsize ConHa};
    \node[fill=black!3, rotate=90, font=\sffamily, minimum width=1.69cm, minimum height=0.45cm] (input) at (-0.203\linewidth,-6.72)  {\scriptsize 0.169bpp};
    \node[fill=black!3, rotate=90, font=\sffamily, minimum width=1.69cm, minimum height=0.45cm] (input) at (-0.235\linewidth,-8.40) {\scriptsize HiFiC};
    \node[fill=black!3, rotate=90, font=\sffamily, minimum width=1.69cm, minimum height=0.45cm] (input) at (-0.203\linewidth,-8.40) {\scriptsize 0.171bpp};
    
    \draw [white, line width=1pt] (0.25\linewidth,3.1) -- (0.25\linewidth,-2.55);
    
    \draw [white, line width=1pt] (0.083333\linewidth,-2.55) -- (0.083333\linewidth,-9.25);
    \draw [white, line width=1pt] (0.416666\linewidth,-2.55) -- (0.416666\linewidth,-9.25);
    
    \draw [white, line width=1pt] (-0.25\linewidth,-2.55) -- (0.75\linewidth,-2.55);
    \draw [white, line width=1pt] (-0.25\linewidth,-4.25) -- (0.75\linewidth,-4.25);
    \draw [white, line width=1pt] (-0.25\linewidth,-5.88) -- (0.75\linewidth,-5.88);
    \draw [white, line width=1pt] (-0.25\linewidth,-7.58) -- (0.75\linewidth,-7.58);

    \end{tikzpicture}
    \caption{Comparison of compression methods. 
    Our approach represents a middle ground between the Hyperprior and HiFiC models.
    For images with pavement (first column), our model adds details to create in-distribution samples.
    It avoids excessive hallucination for images with small faces and text (middle column) or objects with straight edges (right column).
    }
    \label{fig:qual}
\end{figure}

\textbf{Baselines:} 
Our approach integrates the Mean \& Scale Hyperprior \cite{balle2018variational} and HiFiC \cite{mentzer2020high} compression models, both sharing the same architecture.
While the Hyperprior model optimizes solely optimizes Mean Squared Error (MSE) as the distortion metric, HiFiC combines MSE with perceptual loss using LPIPS \cite{zhang2018unreasonable} and a GAN discriminator.
We modify the HiFiC model by making the hyperparameter that controls the perception loss weight image conditional.
To assess the impact of this image conditionality, we also train a model with a fixed weight, ConHa-fixed, calculated as the average weight across the entire training dataset.

\subsection{Qualitative assessment}

Our method strikes a balance between optimizing the rate-distortion trade-off, which can result in blurry images with no hallucinations, and optimizing the rate-distortion-perception trade-off, which can produce in-distribution samples but with too many hallucinations.
In Figure \ref{fig:qual}, we compare our compression model to baseline models using an image featuring diverse objects.
For small faces, straight lines, and text on a shirt, our method avoids excessive generating details that could obscure recognition. 
Conversely, it produces in-distribution samples for textures like road surfaces, larger faces, and rust on a bridge.
Depending on the image content, our method generates samples nearly indistinguishable from either the Hyperprior model or HiFiC.
While any small crop may have a similarly performing compression model, models optimizing for a fixed point on the rate-distortion-perception trade-off tend to underperform across many images.
Thus, our approach consistently generates high-quality images regardless of content, achieving a superior balance in the rate-distortion-perception trade-off.

\subsection{Quantitative assessment} \label{sec:results_user_study}

Computational distortion metrics often fail to predict human preferences accurately, highlighting the need for user studies.
In our study, we collected 1531 comparisons from 40 participants, each completing a maximum of 50 comparisons, with a median time of 7.8 seconds per comparison.
Details aboutthe user study can be found in Appendix \ref{app:user_study}.
Figure \ref{fig:user_study} presents the results, showing models on the x-axis and bootstrapped Elo scores on the y-axis across three bitrates: low, medium, and high.
Across all bitrates, our model consistently outperforms HiFiC, the previous state-of-the-art compression model.
At low bitrates, we compare our model to the extremes of the compression spectrum, Hyperprior and HiFiC, as well as our model with a fixed ratio.
Our method, a middle ground between Hyperprior and HiFiC outperforms both extremes.
When the weight of perceptual losses is fixed instead of being image conditional, performance drops to near HiFiC levels.
This result underscores the importance of conditioning the weight of perceptual losses on the image for optimal performance in image compression.

In Appendix \ref{app:further_results}, we provide a comprehensive comparison of our method and the baseline models using various computational metrics on the  CLIC 2024, CLIC 2020 \cite{compressioncc} and Kodak dataset \cite{kodakgraphics}.
ConHa demonstrates comparable performance to Hyperprior and HiFiC across all metrics, with no definitive best model.
As observed in the CLIC \cite{compressioncc} challenges, automated metrics do not align with human evaluations. 

\definecolor{color1}{HTML}{DD2680}
\definecolor{color2}{HTML}{FE6100}
\definecolor{color3}{HTML}{775EF0}
\definecolor{color4}{HTML}{648FFF}

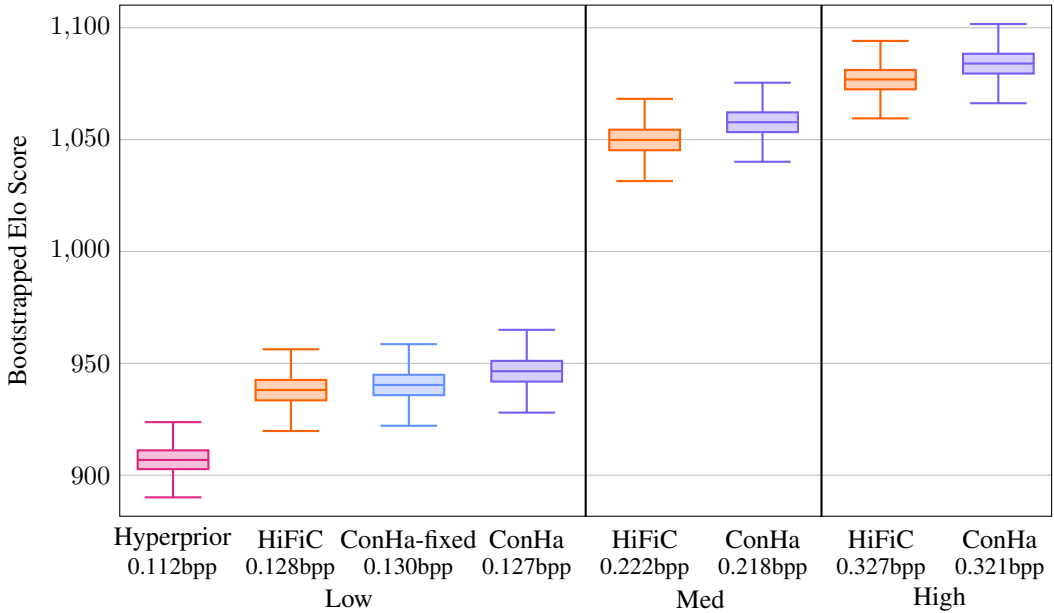
\begin{figure}[!htb]
    \centering
    \begin{tikzpicture}
        \begin{axis}[
            width=\textwidth,
            height=0.6\textwidth,
            boxplot/draw direction=y,
            ylabel={Bootstrapped Elo Score},
            xtick={1, 2, 3, 4, 5, 6, 7, 8},
            xticklabels={, , , , , , , },
            ymajorgrids=true,
            boxplot={
                draw position={1.5},
                box extend=0.6,
            },
            enlarge y limits=0.04,
            enlarge x limits=0.02,
            xtick style={draw=none}, 
            ytick style={draw=none}
        ]

        \addplot+[
            thick,
            color1,
            fill=color1!30,
            solid,
            boxplot prepared={
                lower whisker=890.0915753644388,
                lower quartile=902.6949980910571,
                median=906.805125904239,
                upper quartile=911.0972799088026,
                upper whisker=923.7007026354211,
            },
            boxplot prepared={
                draw position=1
            },
        ] coordinates {};
        
        \addplot+[
            thick,
            color2,
            fill=color2!30,
            solid,
            boxplot prepared={
                lower whisker=919.7461311726067,
                lower quartile=933.4434151739292,
                median=938.0642040739774,
                upper quartile=942.5749378414775,
                upper whisker=956.2722218428,
            },
            boxplot prepared={
                draw position=2
            }
        ] coordinates {};

        \addplot+[
            thick,
            color4,
            fill=color4!30,
            solid,
            boxplot prepared={
                lower whisker=922.0837461249075,
                lower quartile=935.7619249519885,
                median=940.3205740771707,
                upper quartile=944.8807108367091,
                upper whisker=958.5588896637901,
            },
            boxplot prepared={
                draw position=3
            }
        ] coordinates {};

        \addplot+[
            thick,
            color3,
            fill=color3!30,
            solid,
            boxplot prepared={
                lower whisker=927.9543146585673,
                lower quartile=941.835199206268,
                median=946.4526949487586,
                upper quartile=951.0891222380685,
                upper whisker=964.9700067857692,
            },
            boxplot prepared={
                draw position=4
            }
        ] coordinates {};

        \addplot+[
            thick,
            color2,
            fill=color2!30,
            solid,
            boxplot prepared={
                lower whisker=1031.462998072642,
                lower quartile=1045.2493833962535,
                median=1049.8506985004788,
                upper quartile=1054.4403069453278,
                upper whisker=1068.2266922689394,
            },
            boxplot prepared={
                draw position=5
            }
        ] coordinates {};

        \addplot+[
            thick,
            color3,
            fill=color3!30,
            solid,
            boxplot prepared={
                lower whisker=1040.0985632398606,
                lower quartile=1053.3511532624327,
                median=1057.7695738898315,
                upper quartile=1062.1862132774809,
                upper whisker=1075.4388033000532,
            },
            boxplot prepared={
                draw position=6
            }
        ] coordinates {};

        \addplot+[
            thick,
            color2,
            fill=color2!30,
            solid,
            boxplot prepared={
                lower whisker=1059.4901913872477,
                lower quartile=1072.4731159899288,
                median=1076.8975808592072,
                upper quartile=1081.128399058383,
                upper whisker=1094.111323661064,
            },
            boxplot prepared={
                draw position=7
            }
        ] coordinates {};

        \addplot+[
            thick,
            color3,
            fill=color3!30,
            solid,
            boxplot prepared={
                lower whisker=1066.257983022432,
                lower quartile=1079.5327727850536,
                median=1084.0253268295423,
                upper quartile=1088.3826326268013,
                upper whisker=1101.657422389423,
            },
            boxplot prepared={
                draw position=8
            }
        ] coordinates {};

        \draw[thick] (axis cs:4.5,860) -- (axis cs:4.5,1130);
        \draw[thick] (axis cs:6.5,860) -- (axis cs:6.5,1130);        

        \end{axis}
        \node (input) at (3.05, -1.1) {Low};
        \node (input) at (7.72, -1.1) {Med};
        \node (input) at (10.9, -1.1) {High};

        \node (input) at (0.7, -0.3) {Hyperprior};
        \node (input) at (0.7, -0.7) {\small 0.112bpp};
        \node (input) at (2.27, -0.3) {HiFiC};
        \node (input) at (2.27, -0.7) {\small 0.128bpp};
        \node (input) at (3.84, -0.3) {ConHa-fixed};
        \node (input) at (3.84, -0.7) {\small 0.130bpp};
        \node (input) at (5.41, -0.3) {ConHa};
        \node (input) at (5.41, -0.7) {\small 0.127bpp};
        \node (input) at (6.98, -0.3) {HiFiC};
        \node (input) at (6.98, -0.7) {\small 0.222bpp};
        \node (input) at (8.55, -0.3) {ConHa};
        \node (input) at (8.55, -0.7) {\small 0.218bpp};
        \node (input) at (10.12, -0.3) {HiFiC};
        \node (input) at (10.12, -0.7) {\small 0.327bpp};
        \node (input) at (11.7, -0.3) {ConHa};
        \node (input) at (11.7, -0.7) {\small 0.321bpp};


        \node (input) at (0, -1) {};
    \end{tikzpicture}
    \caption{Bootstrapped Elo Scores box plot on the CLIC 2024 image test set. 
    Each box represents the distribution of Elo Scores, with the horizontal line indicating the median, the box extending from the first quartile (Q1) to the third quartile (Q3), and the whiskers extending to 1.5 times the interquartile range (Q3-Q1).
    At low bitrates, ConHa (ours) is compared against three baselines: Hyperprior, HiFiC, and ConHa-fixed.
    For medium and high bitrates, ConHa is compared solely against the previous state-of-the-art, HiFiC.
    Remarkably, at all bitrates, ConHa consistently outperforms the baseline models, as evidenced by the higher median Elo Scores.
    }
    \label{fig:user_study}
\end{figure}

\section{Conclusion}

In the realm of lossy image compression, striking a balance between minimizing distortion and maintaining perceptual fidelity poses a significant challenge. 
Traditional distortion metrics often fail to align with human perceptual preferences, resulting in compression models that either hallucinate excessive details or generate out-of-distribution samples.
By incorporating a classifier trained to predict user preferences regarding detail hallucination, we introduce automatic balancing between distortion and perception in image compression.
This novel approach allows our model to adapt its compression strategy dynamically, optimizing performance across a wide range of image content and compression rates.
ConHa carefully chooses when and what parts of images to hallucinate in a way that aligns with human perception.
By addressing the limitations of distortion metrics and introducing dynamic content-aware balancing between distortion and realism, our method outperforms state-of-the-art compression models.

\FloatBarrier
\newpage
\bibliography{neurips_2024}

\begin{thebibliography}{10}

\bibitem{mentzer2020high}
Fabian Mentzer, George~D Toderici, Michael Tschannen, and Eirikur Agustsson.
\newblock High-fidelity generative image compression.
\newblock {\em Advances in Neural Information Processing Systems},
  33:11913--11924, 2020.

\bibitem{he2022po}
Dailan He, Ziming Yang, Hongjiu Yu, Tongda Xu, Jixiang Luo, Yuan Chen, Chenjian
  Gao, Xinjie Shi, Hongwei Qin, and Yan Wang.
\newblock Po-elic: Perception-oriented efficient learned image coding.
\newblock In {\em Proceedings of the IEEE/CVF Conference on Computer Vision and
  Pattern Recognition}, pages 1764--1769, 2022.

\bibitem{wallace1991jpeg}
Gregory~K Wallace.
\newblock The jpeg still picture compression standard.
\newblock {\em Communications of the ACM}, 34(4):30--44, 1991.

\bibitem{taubman2002jpeg2000}
David~S Taubman and Michael~W Marcellin.
\newblock Jpeg2000: Standard for interactive imaging.
\newblock {\em Proceedings of the IEEE}, 90(8):1336--1357, 2002.

\bibitem{googlewebp}
{Google Developers}.
\newblock Webp.
\newblock \url{https://developers.google.com/speed/webp/}.
\newblock Accessed: May 17, 2024.

\bibitem{bpg}
Fabrice Bellard.
\newblock {BPG (Better Portable Graphics)}.
\newblock \url{https://bellard.org/bpg/}.
\newblock Accessed: May 17, 2024.

\bibitem{kingma2013auto}
Diederik~P Kingma and Max Welling.
\newblock Auto-encoding variational bayes.
\newblock {\em arXiv preprint arXiv:1312.6114}, 2013.

\bibitem{rezende2014stochastic}
Danilo~Jimenez Rezende, Shakir Mohamed, and Daan Wierstra.
\newblock Stochastic backpropagation and approximate inference in deep
  generative models.
\newblock In {\em International conference on machine learning}, pages
  1278--1286. PMLR, 2014.

\bibitem{balle2016end}
Johannes Ball{\'e}, Valero Laparra, and Eero~P Simoncelli.
\newblock End-to-end optimized image compression.
\newblock {\em arXiv preprint arXiv:1611.01704}, 2016.

\bibitem{balle2018variational}
Johannes Ball{\'e}, David Minnen, Saurabh Singh, Sung~Jin Hwang, and Nick
  Johnston.
\newblock Variational image compression with a scale hyperprior.
\newblock {\em arXiv preprint arXiv:1802.01436}, 2018.

\bibitem{he2022elic}
Dailan He, Ziming Yang, Weikun Peng, Rui Ma, Hongwei Qin, and Yan Wang.
\newblock Elic: Efficient learned image compression with unevenly grouped
  space-channel contextual adaptive coding.
\newblock In {\em Proceedings of the IEEE/CVF Conference on Computer Vision and
  Pattern Recognition}, pages 5718--5727, 2022.

\bibitem{compressioncc}
{Compression.cc}.
\newblock {Compression.cc}.
\newblock \url{https://www.compression.cc/}.
\newblock Accessed on May 17, 2024.

\bibitem{zhang2018unreasonable}
Richard Zhang, Phillip Isola, Alexei~A Efros, Eli Shechtman, and Oliver Wang.
\newblock The unreasonable effectiveness of deep features as a perceptual
  metric.
\newblock In {\em Proceedings of the IEEE conference on computer vision and
  pattern recognition}, pages 586--595, 2018.

\bibitem{heusel2017gans}
Martin Heusel, Hubert Ramsauer, Thomas Unterthiner, Bernhard Nessler, and Sepp
  Hochreiter.
\newblock Gans trained by a two time-scale update rule converge to a local nash
  equilibrium.
\newblock {\em Advances in neural information processing systems}, 30, 2017.

\bibitem{goodfellow2014generative}
Ian Goodfellow, Jean Pouget-Abadie, Mehdi Mirza, Bing Xu, David Warde-Farley,
  Sherjil Ozair, Aaron Courville, and Yoshua Bengio.
\newblock Generative adversarial nets.
\newblock {\em Advances in neural information processing systems}, 27, 2014.

\bibitem{rippel2017real}
Oren Rippel and Lubomir Bourdev.
\newblock Real-time adaptive image compression.
\newblock In {\em International Conference on Machine Learning}, pages
  2922--2930. PMLR, 2017.

\bibitem{santurkar2018generative}
Shibani Santurkar, David Budden, and Nir Shavit.
\newblock Generative compression.
\newblock In {\em 2018 Picture Coding Symposium (PCS)}, pages 258--262. IEEE,
  2018.

\bibitem{tschannen2018deep}
Michael Tschannen, Eirikur Agustsson, and Mario Lucic.
\newblock Deep generative models for distribution-preserving lossy compression.
\newblock {\em Advances in neural information processing systems}, 31, 2018.

\bibitem{agustsson2019generative}
Eirikur Agustsson, Michael Tschannen, Fabian Mentzer, Radu Timofte, and Luc~Van
  Gool.
\newblock Generative adversarial networks for extreme learned image
  compression.
\newblock In {\em Proceedings of the IEEE/CVF International Conference on
  Computer Vision}, pages 221--231, 2019.

\bibitem{blau2019rethinking}
Yochai Blau and Tomer Michaeli.
\newblock Rethinking lossy compression: The rate-distortion-perception
  tradeoff.
\newblock In {\em International Conference on Machine Learning}, pages
  675--685. PMLR, 2019.

\bibitem{zhang2021universal}
George Zhang, Jingjing Qian, Jun Chen, and Ashish Khisti.
\newblock Universal rate-distortion-perception representations for lossy
  compression.
\newblock {\em Advances in Neural Information Processing Systems},
  34:11517--11529, 2021.

\bibitem{agustsson2023multi}
Eirikur Agustsson, David Minnen, George Toderici, and Fabian Mentzer.
\newblock Multi-realism image compression with a conditional generator.
\newblock In {\em Proceedings of the IEEE/CVF Conference on Computer Vision and
  Pattern Recognition}, pages 22324--22333, 2023.

\bibitem{iwai2024controlling}
Shoma Iwai, Tomo Miyazaki, and Shinichiro Omachi.
\newblock Controlling rate, distortion, and realism: Towards a single
  comprehensive neural image compression model.
\newblock In {\em Proceedings of the IEEE/CVF Winter Conference on Applications
  of Computer Vision}, pages 2900--2909, 2024.

\bibitem{div2k}
Eirikur Agustsson and Radu Timofte.
\newblock Div2k dataset, 2017.
\newblock Accessed: 2024-05-22.

\bibitem{TOFlow}
Tianfan Xue, Jiajun Wu, Katherine~L. Bouman, and William~T. Freeman.
\newblock {TOFlow}: Temporal output learning for video frame interpolation.
\newblock \url{http://toflow.csail.mit.edu/}, 2019.

\bibitem{kodakgraphics}
Kodak graphics.
\newblock \url{http://r0k.us/graphics/kodak/}.

\bibitem{glickman1995comprehensive}
Mark~E Glickman.
\newblock A comprehensive guide to chess ratings.
\newblock {\em American Chess Journal}, 3(1):59--102, 1995.

\end{thebibliography}

\FloatBarrier
\newpage
\appendix

\section{User study} \label{app:user_study}

\begin{figure}[!ht]
    \centering
    \begin{tikzpicture}
        \node at (0,0) {\includegraphics[width=\textwidth]{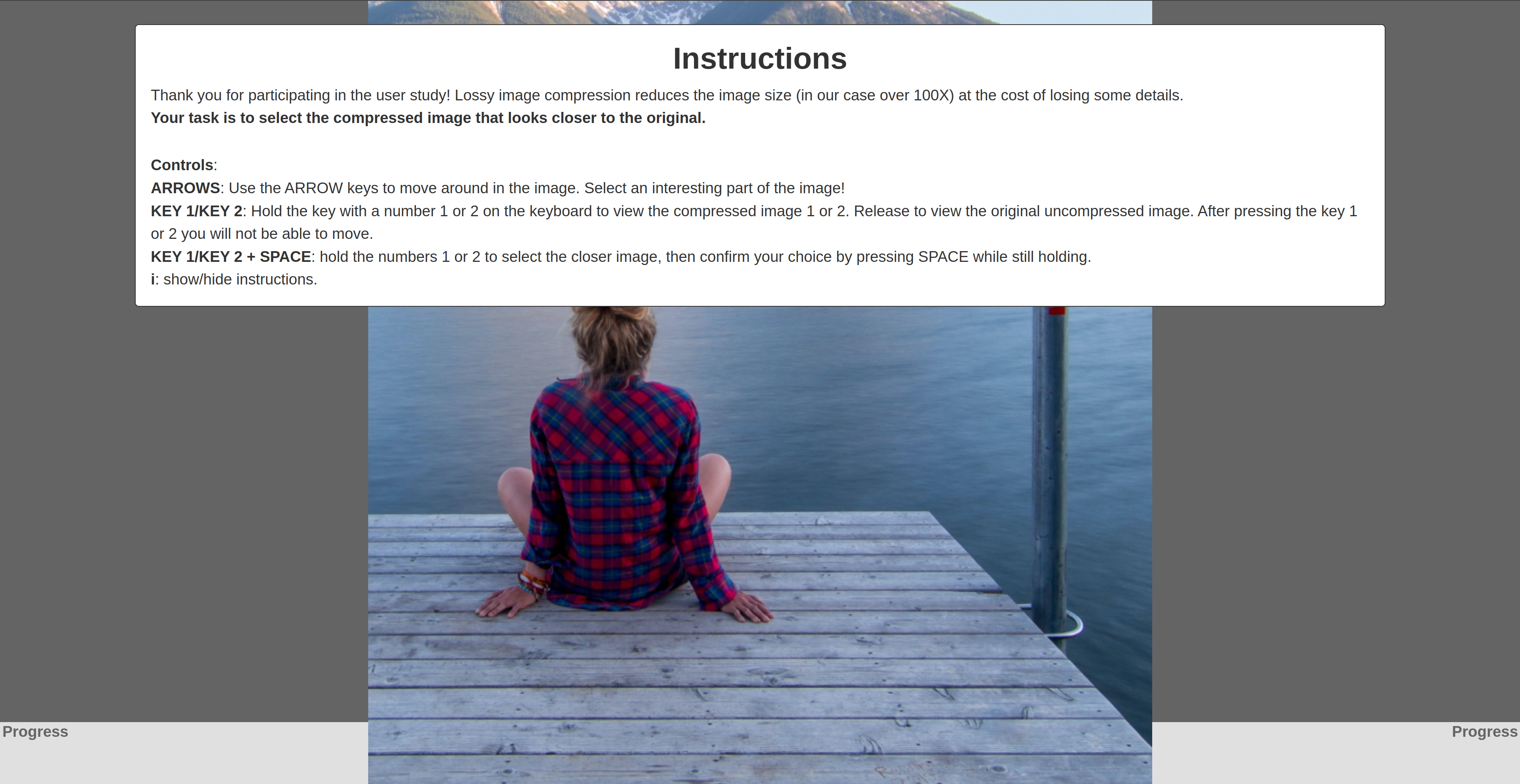}};
        \node at (0,-8) {\includegraphics[width=\textwidth]{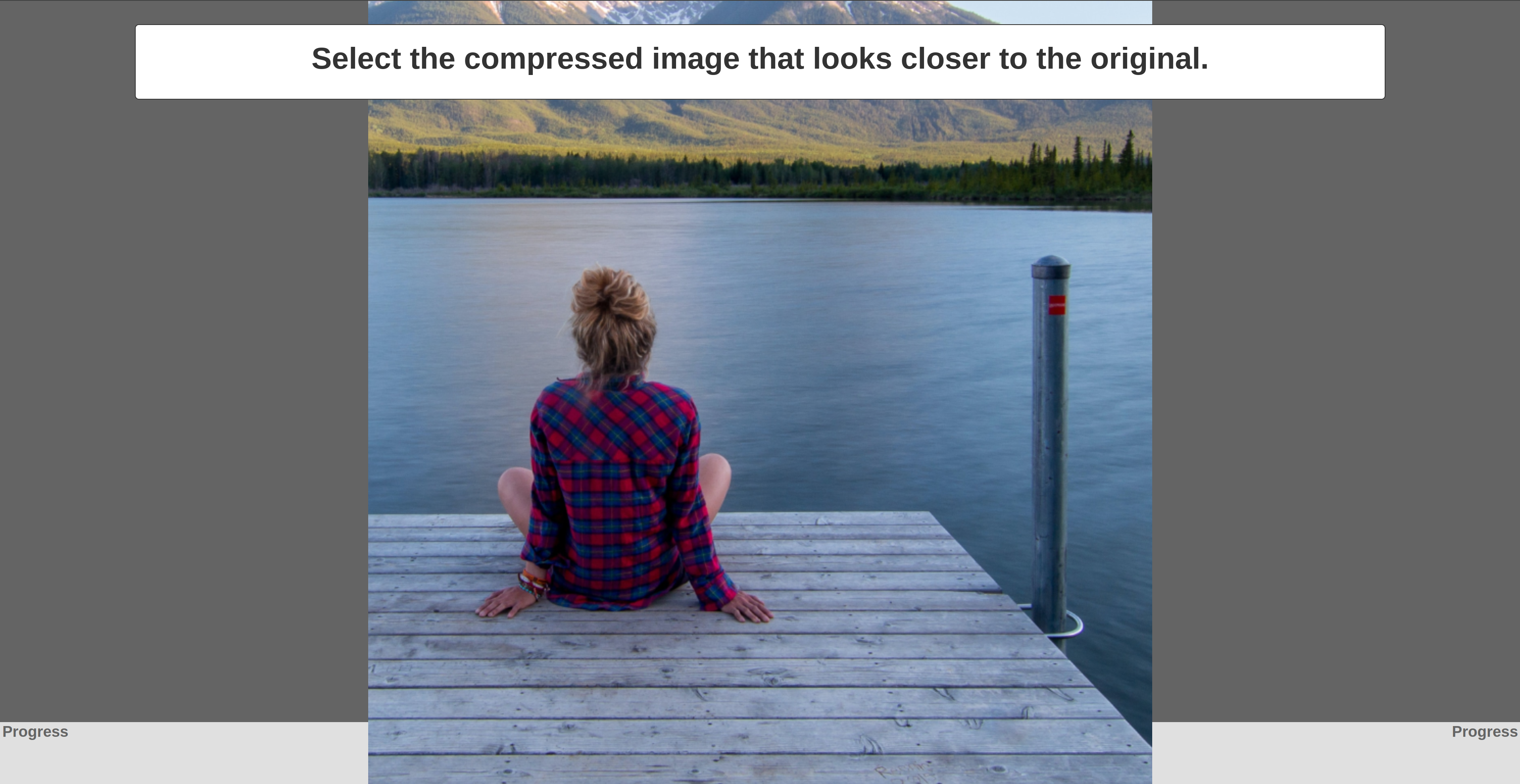}};
    \end{tikzpicture}
    \caption{User study interface with instructions (top) and after closing the instructions (bottom).}
    \label{fig:us_interface}
\end{figure}

The user study is used in both data collectoin for the preference model training and as evaluation.
Following previous studies \cite{compressioncc, mentzer2020high}, we utilize the two-alternative forced choice (2AFC) labeling method.
We decided to implement a web application for our user study, as shown in the screenshot in Figure \ref{fig:us_interface}.
Users begin by seeing the task, ``Select the compressed image that looks closer to the original.'' followed by instructions.
The participant can see the original image in the center of the interface.
Using the arrow keys, they can select an area of interest.
By pressing and holding keys 1 and 2, participants can switch between an image compressed by Model A and Model B. 
To restrict the decision area to 786×786 pixels, panning is disabled once a compressed image is displayed.
To submit their decision, participants hold down either key 1 or 2 and then press the space bar.
The instructions are displayed line by line to avoid overwhelming the participant and can be toggled on and off using the 'I' key. 
The task ``Select the compressed image that looks closer to the original'' is always displayed on the screen. 
Additionally to the label, the interface tracks the number of key presses and the time spent on each image.
The source code is provided in the supplementary material.

\subsection{Data collection} \label{app:data_collection}

To align a compression model with human preferences, training data is essential.
For training data, we compressed images using only two models: the rate-distortion optimized model (Hyperprior) and the rate-distortion-perception optimized model (HiFiC).
The key difference  between these two models is that Hyperprior focuses solely on rate-distortion without a perceptual loss term, while HiFiC incorporates perceptual losses.
As a result, the labels reflect user preferences regarding the trade-off between exact reconstruction with artifacts and hallucinations in in-distribution samples.
Label collection occurred at the 'low' bitrate defined in the HiFiC paper, yielding 5,408 comparisons for the DIV2K training and validation set.
Labels were gathered through crowd-sourcing with volunteers, and these comparisons are available in the supplementary material.

\subsection{Evaluation} \label{app:user_study_evaluation}

During the evaluation phase for each comparison, a 786$\times$78 crop of the original image is presented alongside two compressed images created using different compression models.
To aggregate binary comparisons into model rankings, we use the Elo ranking system \cite{glickman1995comprehensive}.
However, directly applying the Elo algorithm is challenging due to its permutation invariance.
To overcome this, we implement bootstrapping, sampling 10,000 times with replacement and reporting statistics from the final score distribution.

\section{Further results} \label{app:further_results}

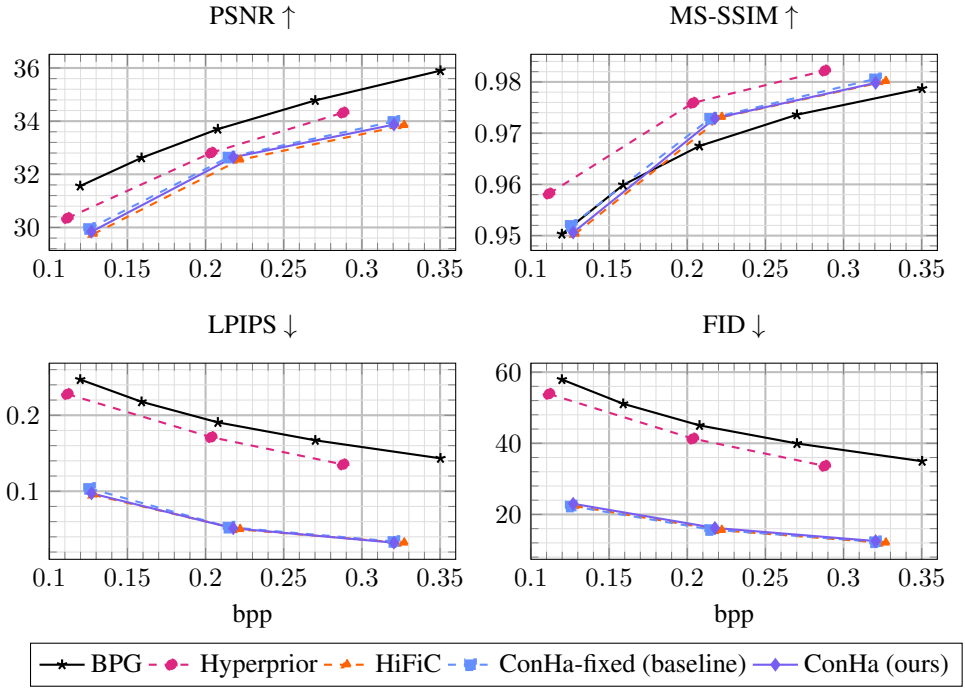
\begin{figure}[!b]
    \centering
    \begin{tikzpicture}
    \begin{groupplot}[
        group style={
            group size=2 by 2,
            xlabels at=edge bottom,
            ylabels at=edge left,
            horizontal sep=1cm,
            vertical sep=1.5cm,
        },
        xlabel={bpp},
        width=0.5\textwidth,
        height=0.3\textwidth,
        grid=both,
        minor tick num=4,
        major grid style={gray!50, thick}, 
        minor grid style={gray!25}, 
        legend style={at={(1.1,-2)},anchor=north},
        legend columns=5,
        xmax=0.36,
        xmin=0.1,
    ]
    
    \nextgroupplot[
        title={PSNR $\uparrow$}
    ]
    \addplot[thick, black, mark=star] coordinates {(0.11981074767,31.56) (0.15904109435,32.62) (0.20785901673,33.7) (0.27018096743,34.78) (0.3502608982,35.9)};
    \addplot[thick, dashed, color=color1, mark=*] coordinates {(0.11158490604,30.35) (0.20362953321,32.81) (0.28799348187,34.32)};
    \addplot[thick, dashed, color=color2, mark=triangle*] coordinates {(0.12821933362,29.75) (0.22179200663,32.54) (0.32675044494,33.84)};
    \addplot[thick, dashed, color=color4, mark=square*] coordinates {(0.12576781504,29.96) (0.21495534682,32.65) (0.32046057345,33.99)};
    \addplot[thick, color=color3, mark=diamond*] coordinates {(0.1270257171,29.84) (0.21769759499,32.64) (0.32051944168,33.87)};
    \addlegendentry{BPG}
    \addlegendentry{Hyperprior}
    \addlegendentry{HiFiC}
    \addlegendentry{ConHa-fixed (baseline)}
    \addlegendentry{ConHa (ours)}
    
    \nextgroupplot[
        title={MS-SSIM $\uparrow$}
    ]
    \addplot[thick, black, mark=star] coordinates {(0.11981074767,0.9503) (0.15904109435,0.9599) (0.20785901673,0.9675) (0.27018096743,0.9736) (0.3502608982,0.9787)};
    \addplot[thick, dashed, color=color1, mark=*] coordinates {(0.11158490604,0.9582) (0.20362953321,0.9759) (0.28799348187,0.9822)};
    \addplot[thick, dashed, color=color2, mark=triangle*] coordinates {(0.12821933362,0.9503) (0.22179200663,0.9731) (0.32675044494,0.9801)};
    \addplot[thick, dashed, color=color4, mark=square*] coordinates {(0.12576781504,0.952) (0.21495534682,0.9729) (0.32046057345,0.9806)};
    \addplot[thick, color=color3, mark=diamond*] coordinates {(0.1270257171,0.9506) (0.21769759499,0.9729) (0.32051944168,0.9798)};
    
    \nextgroupplot[
        title={LPIPS $\downarrow$}
    ]
    \addplot[thick, black, mark=star] coordinates {(0.11981074767,0.247) (0.15904109435,0.2177) (0.20785901673,0.1906) (0.27018096743,0.167) (0.3502608982,0.1433)};
    \addplot[thick, dashed, color=color1, mark=*] coordinates {(0.11158490604,0.2278) (0.20362953321,0.1711) (0.28799348187,0.1352)};
    \addplot[thick, dashed, color=color2, mark=triangle*] coordinates {(0.12821933362,0.09527) (0.22179200663,0.04974) (0.32675044494,0.03193)};
    \addplot[thick, dashed, color=color4, mark=square*] coordinates {(0.12576781504,0.1037) (0.21495534682,0.05273) (0.32046057345,0.03389)};
    \addplot[thick, color=color3, mark=diamond*] coordinates {(0.1270257171,0.09717) (0.21769759499,0.05125) (0.32051944168,0.0323)};
    
    \nextgroupplot[
        title={FID $\downarrow$}
    ]
    \addplot[thick, black, mark=star] coordinates {(0.11981074767,57.92) (0.15904109435,51.1) (0.20785901673,45.03) (0.27018096743,39.94) (0.3502608982,34.95) };
    \addplot[thick, dashed, color=color1, mark=*] coordinates {(0.11158490604,53.86) (0.20362953321,41.3) (0.28799348187,33.64)};
    \addplot[thick, dashed, color=color2, mark=triangle*] coordinates {(0.12821933362,22.33) (0.22179200663,15.53) (0.32675044494,11.98)};
    \addplot[thick, dashed, color=color4, mark=square*] coordinates {(0.12576781504,22.35) (0.21495534682,15.76) (0.32046057345,12.32)};
    \addplot[thick, color=color3, mark=diamond*] coordinates {(0.1270257171,22.99) (0.21769759499,16.2) (0.32051944168,12.51)};
    
    \end{groupplot}
    \end{tikzpicture}
    \caption{Rate-distortion and -perception curves on CLIC 2024 image test set. Arrows indicate whether higher ($\uparrow$) or lower ($\downarrow$) values are preferable.}
    \label{fig:evaluation_metrics}
\end{figure}

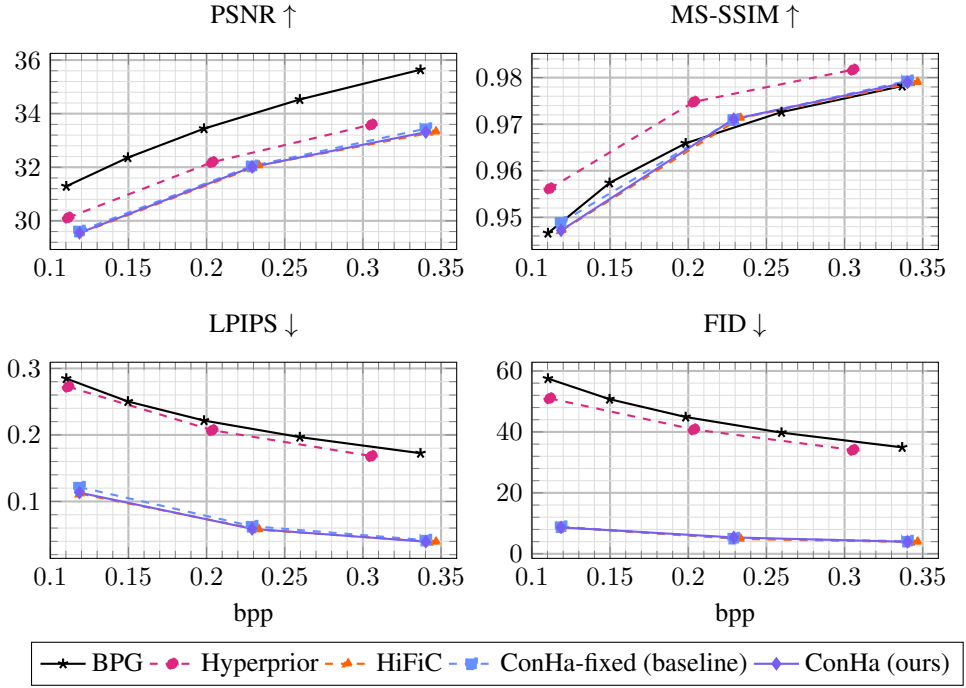
\begin{figure}[!htb]
    \centering
    \begin{tikzpicture}
    \begin{groupplot}[
        group style={
            group size=2 by 2,
            xlabels at=edge bottom,
            ylabels at=edge left,
            horizontal sep=1cm,
            vertical sep=1.5cm,
        },
        xlabel={bpp},
        width=0.5\textwidth,
        height=0.3\textwidth,
        grid=both,
        minor tick num=4,
        major grid style={gray!50, thick}, 
        minor grid style={gray!25}, 
        legend style={at={(1.1,-2)},anchor=north},
        legend columns=5,
        xmax=0.36,
        xmin=0.1,
    ]
    
    \nextgroupplot[
        title={PSNR $\uparrow$}
    ]
    \addplot[thick, black, mark=star] coordinates {(0.11025951201,31.29) (0.1496671832,32.36) (0.19835325836,33.44) (0.25966415393,34.53) (0.33685886355,35.64)};
    \addplot[thick, dashed, color=color1, mark=*] coordinates {(0.11145986044,30.12) (0.20362953321,32.19) (0.30547525833,33.59)};
    \addplot[thick, dashed, color=color2, mark=triangle*] coordinates {(0.11925972951,29.54) (0.23321166917,32.06) (0.34644370322,33.32)};
    \addplot[thick, dashed, color=color4, mark=square*] coordinates {(0.11868819713,29.62) (0.22913409403,32.05) (0.34026955109,33.44)};
    \addplot[thick, color=color3, mark=diamond*] coordinates {(0.11868819713,29.55) (0.22913409403,32.01) (0.34026955109,33.31)};
    \addlegendentry{BPG}
    \addlegendentry{Hyperprior}
    \addlegendentry{HiFiC}
    \addlegendentry{ConHa-fixed (baseline)}
    \addlegendentry{ConHa (ours)}
    
    \nextgroupplot[
        title={MS-SSIM $\uparrow$}
    ]
    \addplot[thick, black, mark=star] coordinates {(0.11025951201,0.9466) (0.1496671832,0.9574) (0.19835325836,0.9659) (0.25966415393,0.9726) (0.33685886355,0.9782)};
    \addplot[thick, dashed, color=color1, mark=*] coordinates {(0.11145986044,0.9562) (0.20362953321,0.9748) (0.30547525833,0.9817)};
    \addplot[thick, dashed, color=color2, mark=triangle*] coordinates {(0.11925972951,0.9472) (0.23321166917,0.9713) (0.34644370322,0.979)};
    \addplot[thick, dashed, color=color4, mark=square*] coordinates {(0.11868819713,0.9489) (0.22913409403,0.971) (0.34026955109,0.9793)};
    \addplot[thick, color=color3, mark=diamond*] coordinates {(0.11868819713,0.9472) (0.22913409403,0.9711) (0.34026955109,0.979)};
    
    \nextgroupplot[
        title={LPIPS $\downarrow$}
    ]
    \addplot[thick, black, mark=star] coordinates {(0.11025951201,0.2847) (0.1496671832,0.2503) (0.19835325836,0.2216) (0.25966415393,0.1967) (0.33685886355,0.1725)};
    \addplot[thick, dashed, color=color1, mark=*] coordinates {(0.11145986044,0.2724) (0.20362953321,0.2071) (0.30547525833,0.1682)};
    \addplot[thick, dashed, color=color2, mark=triangle*] coordinates {(0.11925972951,0.1115) (0.23321166917,0.05767) (0.34644370322,0.039)};
    \addplot[thick, dashed, color=color4, mark=square*] coordinates {(0.11868819713,0.1215) (0.22913409403,0.06277) (0.34026955109,0.04185)};
    \addplot[thick, color=color3, mark=diamond*] coordinates {(0.11868819713,0.1133) (0.22913409403,0.05882) (0.34026955109,0.03955)};
    
    \nextgroupplot[
        title={FID $\downarrow$}
    ]
    \addplot[thick, black, mark=star] coordinates {(0.11025951201,57.5) (0.1496671832,50.74) (0.19835325836,44.89) (0.25966415393,39.76) (0.33685886355,34.93) };
    \addplot[thick, dashed, color=color1, mark=*] coordinates {(0.11145986044,51.0) (0.20362953321,40.8) (0.30547525833,34.07)};
    \addplot[thick, dashed, color=color2, mark=triangle*] coordinates {(0.11925972951,8.672) (0.23321166917,4.846) (0.34644370322,3.867)};
    \addplot[thick, dashed, color=color4, mark=square*] coordinates {(0.11868819713,8.883) (0.22913409403,5.137) (0.34026955109,4.147)};
    \addplot[thick, color=color3, mark=diamond*] coordinates {(0.11868819713,8.672) (0.22913409403,5.373) (0.34026955109,3.938)};
    
    \end{groupplot}
    \end{tikzpicture}
    \caption{Rate-distortion and -perception curves on CLIC 2020 test set. Arrows indicate whether higher ($\uparrow$) or lower ($\downarrow$) values are preferable.}
    \label{fig:a1_evaluation_metrics}
\end{figure}

\begin{figure}[!htb]
    \centering
    \begin{tikzpicture}
    \begin{groupplot}[
        group style={
            group size=2 by 2,
            xlabels at=edge bottom,
            ylabels at=edge left,
            horizontal sep=1cm,
            vertical sep=1.5cm,
        },
        xlabel={bpp},
        width=0.5\textwidth,
        height=0.3\textwidth,
        grid=both,
        minor tick num=4,
        major grid style={gray!50, thick}, 
        minor grid style={gray!25}, 
        legend style={at={(1.1,-2)},anchor=north},
        legend columns=5,
        xmax=0.6,
        xmin=0.2,
    ]
    
    \nextgroupplot[
        title={PSNR $\uparrow$}
    ]
    \addplot[thick, black, mark=star] coordinates {(0.16059451633,28.67) (0.22322675916,29.8) (0.30298021104,30.97) (0.40539635552,32.23) (0.53275044759,33.54)};
    \addplot[thick, dashed, color=color1, mark=*] coordinates {(0.22582668728,27.7) (0.37919447157,30.08) (0.5215250651,31.72)};
    \addplot[thick, dashed, color=color2, mark=triangle*] coordinates {(0.23275078667,26.92) (0.40257771809,29.68) (0.56620958116,31.28)};
    \addplot[thick, dashed, color=color4, mark=square*] coordinates {(0.23208279079,27.06) (0.39281887478,29.74) (0.56050279405,31.38)};
    \addplot[thick, color=color3, mark=diamond*] coordinates {(0.23179117838,26.92) (0.39677598741,29.67) (0.56023491753,31.21)};
    \addlegendentry{BPG}
    \addlegendentry{Hyperprior}
    \addlegendentry{HiFiC}
    \addlegendentry{ConHa-fixed (baseline)}
    \addlegendentry{ConHa (ours)}
    
    \nextgroupplot[
        title={MS-SSIM $\uparrow$}
    ]
    \addplot[thick, black, mark=star] coordinates {(0.16059451633,0.9241) (0.22322675916,0.9415) (0.30298021104,0.9554) (0.40539635552,0.9661) (0.53275044759,0.9747)};
    \addplot[thick, dashed, color=color1, mark=*] coordinates {(0.22582668728,0.9359) (0.37919447157,0.9659) (0.5215250651,0.977)};
    \addplot[thick, dashed, color=color2, mark=triangle*] coordinates {(0.23275078667,0.9224) (0.40257771809,0.9616) (0.56620958116,0.9743)};
    \addplot[thick, dashed, color=color4, mark=square*] coordinates {(0.23208279079,0.9251) (0.39281887478,0.9612) (0.56050279405,0.9746)};
    \addplot[thick, color=color3, mark=diamond*] coordinates {(0.23179117838,0.9224) (0.39677598741,0.9613) (0.56023491753,0.974)};
    
    \nextgroupplot[
        title={LPIPS $\downarrow$}
    ]
    \addplot[thick, black, mark=star] coordinates {(0.16059451633,0.311) (0.22322675916,0.2599) (0.30298021104,0.213) (0.40539635552,0.1713) (0.53275044759,0.1334)};
    \addplot[thick, dashed, color=color1, mark=*] coordinates {(0.22582668728,0.3) (0.37919447157,0.1945) (0.5215250651,0.1375)};
    \addplot[thick, dashed, color=color2, mark=triangle*] coordinates {(0.23275078667,0.1336) (0.40257771809,0.063) (0.56620958116,0.03809)};
    \addplot[thick, dashed, color=color4, mark=square*] coordinates {(0.23208279079,0.1423) (0.39281887478,0.06878) (0.56050279405,0.04152)};
    \addplot[thick, color=color3, mark=diamond*] coordinates {(0.23179117838,0.1347) (0.39677598741,0.06365) (0.56023491753,0.03915)};
    
    \nextgroupplot[
        title={FID $\downarrow$}
    ]
    \addplot[thick, black, mark=star] coordinates {(0.16059451633,103.8) (0.22322675916,90.31) (0.30298021104,76.43) (0.40539635552,65.43) (0.53275044759,55.11) };
    \addplot[thick, dashed, color=color1, mark=*] coordinates {(0.22582668728,94.29) (0.37919447157,73.31) (0.5215250651,59.17)};
    \addplot[thick, dashed, color=color2, mark=triangle*] coordinates {(0.23275078667,45.32) (0.40257771809,30.32) (0.56620958116,24.75)};
    \addplot[thick, dashed, color=color4, mark=square*] coordinates {(0.23208279079,46.39) (0.39281887478,31.77) (0.56050279405,25.46)};
    \addplot[thick, color=color3, mark=diamond*] coordinates {(0.23179117838,46.42) (0.39677598741,31.16) (0.56023491753,24.75)};
    
    \end{groupplot}
    \end{tikzpicture}
    \caption{Rate-distortion and -perception curves on the Kodak dataset. Arrows indicate whether higher ($\uparrow$) or lower ($\downarrow$) values are preferable.}
    \label{fig:a2_evaluation_metrics}
\end{figure}
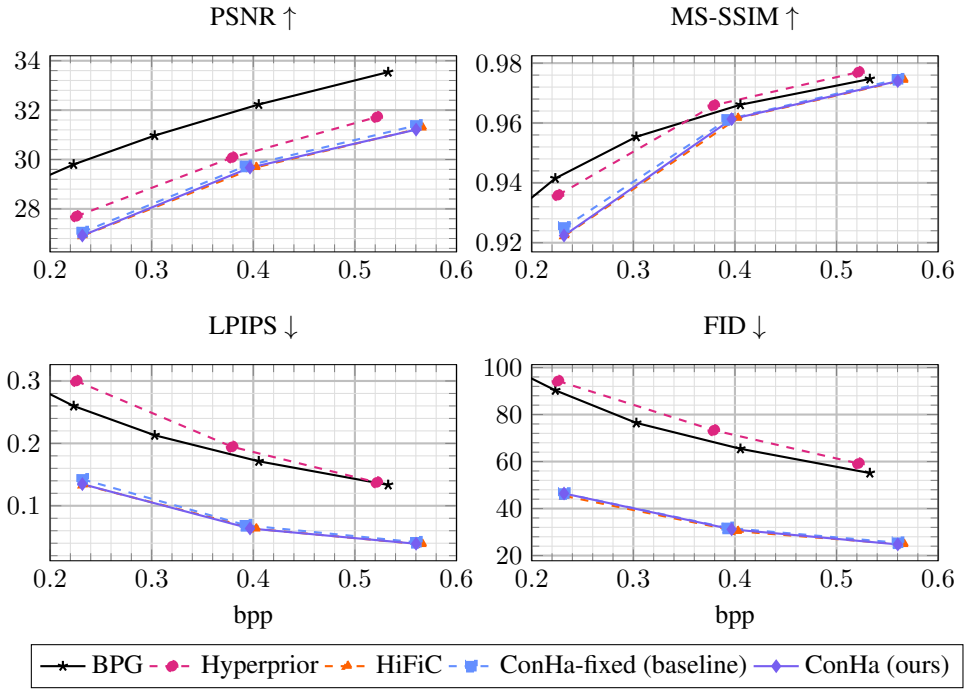

In Figures \ref{fig:evaluation_metrics}, \ref{fig:a1_evaluation_metrics} and \ref{fig:a2_evaluation_metrics}, we provide a comprehensive comparison of our method and the baseline models using various computational metrics.
BPG performs best in terms of PSNR, followed by Hyperprior, with perceptual compression models trailing behind.
However, in terms of MS-SSIM, other methods catch up to BPG.
Notably, GAN-based methods such as HiFiC, ConHa-fixed, and ConHa exhibit lower (thus better) LPIPS and FID scores.
Our method shares similarities with HiFiC, exhibiting slightly higher PSNR, comparable MS-SSIM and LPIPS scores, and marginally worse FID scores.
This suggests that our method achieves a balance between traditional fidelity metrics and perceptual quality metrics.
The shortcomings of the distortion metrics are evident, as they do not capture the performance gap between ConHa and other baselines, as indicated by the user study.
It's worth noting that the FID score quantifies the disparity between the original and distorted distributions.




\section{Training details} \label{app:training_details}

Both models are trained on random 256 $\times$ 256 crops, the preference model on the DIV2K \cite{div2k}, while the image compressor on the Vimeo90k dataset \cite{TOFlow}. Data collection for the preference model training is described in Appendix \ref{app:data_collection}.

\textbf{Preference model:} 

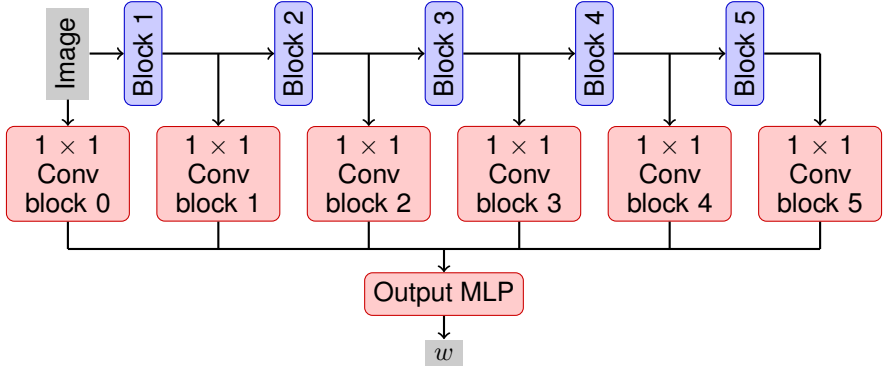
\begin{figure}[!ht]
    \centering
    \begin{tikzpicture}[
        conv/.style={rectangle, draw=blue!80!black, fill=blue!20, minimum width=1cm, minimum height=0.5cm, font=\sffamily, rounded corners=3pt},
        mlp/.style={rectangle, draw=red!80!black, fill=red!20, minimum width=1cm, minimum height=0.5cm, font=\sffamily, rounded corners=3pt},
        arrow/.style={->, thick}
    ]
    
    \node[fill=black!20, rotate=90, font=\sffamily] (input) at (1,0) {Image};
    
    \node[conv, rotate=90] (block1) at (2,0) {Block 1};
    \node[conv, rotate=90] (block2) at (4,0) {Block 2};
    \node[conv, rotate=90] (block3) at (6,0) {Block 3};
    \node[conv, rotate=90] (block4) at (8,0) {Block 4};
    \node[conv, rotate=90] (block5) at (10,0) {Block 5};
    
    \node[mlp, text width=1.4cm, align=center] (mlp0) at (1, -1.6) {1 $\times$ 1 Conv block 0};
    \node[mlp, text width=1.4cm, align=center] (mlp1) at (3, -1.6) {1 $\times$ 1 Conv block 1};
    \node[mlp, text width=1.4cm, align=center] (mlp2) at (5, -1.6) {1 $\times$ 1 Conv block 2};
    \node[mlp, text width=1.4cm, align=center] (mlp3) at (7, -1.6) {1 $\times$ 1 Conv block 3};
    \node[mlp, text width=1.4cm, align=center] (mlp4) at (9, -1.6) {1 $\times$ 1 Conv block 4};
    \node[mlp, text width=1.4cm, align=center] (mlp5) at (11, -1.6) {1 $\times$ 1 Conv block 5};
    
    \draw[arrow] (input) -- (block1);
    \draw[arrow] (block1) -- (block2);
    \draw[arrow] (block2) -- (block3);
    \draw[arrow] (block3) -- (block4);
    \draw[arrow] (block4) -- (block5);
    \draw[thick] (block5) -- (11, 0);
    
    \draw[arrow] (input) -- (mlp0);
    \draw[arrow] (3, 0) -- (mlp1);
    \draw[arrow] (5, 0) -- (mlp2);
    \draw[arrow] (7, 0) -- (mlp3);
    \draw[arrow] (9, 0) -- (mlp4);
    \draw[arrow] (11, 0) -- (mlp5);

    \node[mlp] (outputmlp) at (6, -3.2) {Output MLP};

    \draw[thick] (1, -2.6) -- (11, -2.6);
    \draw[thick] (mlp0) -- (1, -2.6);
    \draw[thick] (mlp1) -- (3, -2.6);
    \draw[thick] (mlp2) -- (5, -2.6);
    \draw[thick] (mlp3) -- (7, -2.6);
    \draw[thick] (mlp4) -- (9, -2.6);
    \draw[thick] (mlp5) -- (11, -2.6);
    \draw[arrow] (6, -2.6) -- (outputmlp);

    \node[fill=black!20, font=\sffamily] (output) at (6,-4.0) {$w$};
    \draw[arrow] (outputmlp) -- (output);
    
    \end{tikzpicture}
    \caption{The hallucination-distribution preference model architecture. 
    Features are extracted with a pre-trained frozen ResNet50, then fed into a block of convolutional layers with 1 $\times$ 1 kernel. 
    The 1 $\times$ 1 convolutional block outputs are average pooled, then concatenated and fed into 2 feed-forward layers with ReLU activation in-between.}
    \label{fig:hd_arh}
\end{figure}

The features are extracted from image $x$ using a frozen pre-trained ResNet50, and a small head is trained on these features.
For most images, the $M_{P}$ model predicts weights in the range of [0.5, 0.6].
To push the weight $w$ to closer to 0 or 1 during compression model training, we multiply the logit before the sigmoid function by 100.
The architecture is illustrated in Figure \ref{fig:hd_arh}.
The configuration of layers in the 1$\times$1 Conv blocks is presented in Table \ref{tab:conv_block_arch}. 
We collected labels at the 'low' bitrate, following the definition in the HiFiC paper, resulting in 5408 pairwise comparisons for the DIV2K training and validation dataset. 
Training took 30 epochs, employing a batch size of 8. 
The Adam optimizer is utilized with a learning rate of $1 \times 10^{-4}$ and cosine learning rate annealing. 
We employ the binary cross-entropy loss function.
Training the preference model takes less than an hour on an Nvidia A100 GPU.

\begin{table}[!ht]
    \centering
    \begin{tabular}{c|cccccc}
         layer & block 0 & block 1 & block 2 & block 3 & block 4 & block 5 \\ \midrule \midrule
         1 & 16 & 64 & 64 & 256 & 256 & 256 \\
         2 & 32 & 32 & 32 & 64 & 64 & 64 \\
         3 & 16 & 16 & 16 & 16 & 16 & 16 \\
    \end{tabular}
    \caption{The 1$\times$1 Conv block number of neurons per layer.}
    \label{tab:conv_block_arch}
\end{table}

\textbf{Compression model:} 
We adopt the HiFiC approach \cite{mentzer2020high} for our compression model training, with one notable adjustment regarding the training dataset.
While HiFiC utilized their proprietary dataset, we employ the Vimeo90k dataset \cite{TOFlow} due to accessibility constraints.
As HiFiC we train the M\&S Hyperprior model for 2M iterations and HiFiC, ConHa-fix, and ConHa follow a two-stage training.
First, we train the model for 1 million iterations with the rate, MSE, and LPIPS losses.
Then we continue training for another 1 million iterations with the GAN component incorporated, together the two-stage training takes 2 million steps as well.
Notably, training our compression model for low, medium, or high bitrates is efficient, requiring less than 3 days on an A100 GPU.

\section{Limitations} \label{sec:limitations}

Our work builds on top of rate-distortion and rate-distortion-perception optimized VAEs.
We aim to strike a balance by automatically selecting the most suitable compression method based on image content.
Human preferences between the two compression models may not vary significantly depending on the compressed image. 
Our model closely resembles the rate-distortion-perception compression model, differing only in cases where the state-of-the-art underperforms the rate-distortion optimized model.
Our improvements are noticeable only through user studies, as existing metrics struggle to capture such nuanced human preferences.
Enhancing distortion metrics may enable them to detect subtle differences, but it may also render our work obsolete.
Distortion metrics would then consider visible compression artifacts and whether the image appears aesthetically pleasing.

\end{document}